\documentclass[a4paper,fleqn]{cas-dc}

\usepackage[numbers]{natbib}
\usepackage{float}
\usepackage{amssymb}
\usepackage{amsmath}
\usepackage{fontenc}
\usepackage{float}
\usepackage{multirow}
\usepackage{booktabs}
\usepackage{array}
\usepackage{makecell}

\def\tsc#1{\csdef{#1}{\textsc{\lowercase{#1}}\xspace}}
\tsc{WGM}
\tsc{QE}

\begin{document}
\let\WriteBookmarks\relax
\def\floatpagepagefraction{1}
\def\textpagefraction{.001}

\shorttitle{}    

\shortauthors{}  

\title [mode = title]{Enhancing Security in Deep Reinforcement Learning: A Comprehensive Survey on Adversarial Attacks and Defenses}  



%

\author[au]{Wu Yichao} 
\author[au]{Wang Yirui}
\author[au]{Ding Panpan}
\author[au]{Wang Hailong}
\author[au]{Zhu Bingqian}
\author[au,co]{Liu Chun\textsuperscript{*}} 
\ead{liuchun@henu.edu.cn}   
\cortext[cor1]{Corresponding auther}

\affiliation[au]{organization={School of Computer and Information Engineering, Henan University},
            city={Kaifeng},
            postcode={475001},
            country={China}}

\affiliation[co]{organization={Henan Industrial Technology Academy of Spatio-Temporal Big Data, Henan University},
	city={Zhengzhou},
	postcode={450000},
     country={China}
}




\begin{abstract}
With the wide application of deep reinforcement learning (DRL) techniques in complex fields such as autonomous driving, intelligent manufacturing, and smart healthcare, how to improve its security and robustness in dynamic and changeable environments has become a core issue in current research. Especially in the face of adversarial attacks, DRL may suffer serious performance degradation or even make potentially dangerous decisions, so it is crucial to ensure their stability in security-sensitive scenarios. In this paper, we first introduce the basic framework of DRL and analyze the main security challenges faced in complex and changing environments. In addition, this paper proposes an adversarial attack classification framework based on perturbation type and attack target and reviews the mainstream adversarial attack methods against DRL in detail, including various attack methods such as perturbation state space, action space, reward function and model space. To effectively counter the attacks, this paper systematically summarizes various current robustness training strategies, including adversarial training, competitive training, robust learning, adversarial detection, defense distillation and other related defense techniques, we also discuss the advantages and shortcomings of these methods in improving the robustness of DRL. Finally, this paper looks into the future research direction of DRL in adversarial environments, emphasizing the research needs in terms of improving generalization, reducing computational complexity, and enhancing scalability and explainability, aiming to provide valuable references and directions for researchers.
\end{abstract}


\begin{highlights}
\item Proposes a taxonomy of adversarial attacks on DRL based on perturbation type and target. 
\item Provides a detailed review of state, action, reward, and model space attack methods in DRL. 
\item Constructs a classification of adversarial defenses, covering training, detection, and distillation. 
\item Identifies key challenges in DRL security, including generalization, efficiency, and scalability.  
\item Suggests future directions to improve robustness, adaptability, and reliability in adversarial DRL.  
\end{highlights}


\begin{keywords}
 Deep Reinforcement Learning\sep Adversarial Attacks\sep Robustness\sep Defense Strategies\sep Security Challenges
\end{keywords}

\maketitle











\section{intoduction}
In recent years, sustained national attention and robust support for a wide range of emerging industries, including intelligent manufacturing\cite{zhouji}, autonomous driving\cite{ZGGL}, embodied intelligence\cite{lisongyuan}, and large-scale models\cite{yuguoming}, have catalyzed extensive research in autonomous learning driven by multimodal information processing, aiming to perceive and interpret heterogeneous data from sources such as vision, audio, and text, and to enable intelligent decision-making and dynamic feedback in complex environments by deeply exploring inter-data correlations and high-dimensional feature representations. As a key paradigm of autonomous learning, Reinforcement Learning (RL) continuously seeks optimal action policies through interactions with real-world environments and self-play, with the goal of maximizing cumulative rewards, while DRL further integrates perceptual deep neural networks (DNNs) with decision-making algorithms to enhance agents’ perception, reasoning, and policy optimization in high-dimensional state spaces, achieving remarkable progress in recent years and expanding its applicability across increasingly complex domains. Landmark achievements in strategic games illustrate this progress: AlphaGo\cite{silver2016mastering}, developed by Google DeepMind, combined DRL with Monte Carlo Tree Search (MCTS) to defeat the world champion Lee Sedol in Go, and its successor, AlphaZero\cite{silver2017mastering}, eliminated reliance on human expert data, learning solely through self-play to achieve superhuman performance across Go, chess, and shogi. In multi-agent cooperation and competitive scenarios, OpenAI Five\cite{berner2019dota} and AlphaStar\cite{vinyals2019grandmaster} reached professional and even superhuman levels in Dota 2 and StarCraft II, respectively, by leveraging Proximal Policy Optimization (PPO)\cite{schulman2017proximal}, Recurrent Neural Networks (RNNs)\cite{sherstinsky2020fundamentals}, and large-scale distributed training to realize highly coordinated strategies and adaptive planning, demonstrating DRL’s potent real-time decision-making capabilities in dynamic environments. Beyond games, DRL has advanced autonomous driving and mobile robotics: in autonomous driving\cite{kiran2021deep}, it has enabled end-to-end driving decisions by efficiently processing high-dimensional sensor inputs such as LiDAR and cameras for complex path planning and dynamic obstacle avoidance, whereas in mobile robotics\cite{zhu2021deep}, it has markedly improved autonomous navigation, task planning, and environmental interaction, with goal-driven navigation algorithms allowing agents to perform efficiently in unfamiliar environments, and model-based DRL methods reducing sample complexity and enhancing performance under resource constraints. Collectively, these advances suggest that, with continued research in DRL and related fields, humanity may progressively approach the ambitious vision of “solving intelligence, and then using intelligence to solve everything”\cite{duwei}.

Although RL has achieved groundbreaking progress across various domains, demonstrating strong decision-optimization capabilities in complex dynamic environments, research on its security remains comparatively limited. In recent years, with the emergence of adversarial attack techniques against deep learning models, the robustness and security of RL models have increasingly attracted attention from both academia and industry. In practical applications, RL models must not only cope with uncertainties in dynamic interactive environments within high-dimensional state spaces but also withstand malicious adversarial attacks. Such attacks, often realized through carefully designed perturbations or interventions, can substantially degrade model performance and even mislead agents into potentially hazardous decisions\cite{javed2024robustness}. Typically, adversarial attacks against DRL systems target five key aspects: the environment, observations, rewards, actions, and policies. Therefore, prior to the widespread deployment of DRL systems in industrial and safety-critical domains, systematically investigating their vulnerabilities and exploring methods to enhance robustness has become a crucial research direction to ensure system safety and reliability.

Existing reviews have predominantly focused on the domain of supervised learning, covering the development of adversarial attack and defense methods for classification models, as well as the application of explainable artificial intelligence in adversarial machine learning \cite{buivh2024critical,yamagata2024safe,baniecki2024adversarial}. To the best of our knowledge, there is currently no comprehensive analysis or systematic review specifically addressing adversarial attack methods and adversarial training strategies for DRL models. To fill this research gap, this paper systematically summarizes representative literature published between 2021 and 2025 and innovatively proposes a classification framework for adversarial attack methods based on perturbation types and attack targets. In addition, this work provides a comprehensive review of existing robustness-enhancing training strategies for DRL agents and organizes these methods into coherent categories. The primary objective of this paper is to offer researchers a comprehensive and detailed perspective to better understand the methodologies and current state of research on various adversarial attacks and training strategies, while also providing guidance for the design and optimization of robustness strategies in practical applications.

The main contributions of this paper are as follows:

1. We propose a classification framework for adversarial attack methods against DRL models based on perturbation types and attack targets.

2. Leveraging this framework, we comprehensively review existing adversarial attack methods and their applications.

3. We systematically summarize various training strategies aimed at enhancing the robustness of DRL agents.

4. We outline future research directions for robustness studies of DRL agents in adversarial environments.

The organization of this paper is illustrated in Figure 1. Section 2 introduces the theoretical foundations of RL and DNNs and provides an overview of the associated security challenges. Section 3 presents the proposed classification system for adversarial attack methods based on perturbation types and attack targets. Section 4 reviews and categorizes existing robustness training strategies and discusses their effectiveness in enhancing agent resilience. Section 5 highlights potential future research directions for DRL agent robustness in adversarial environments. Finally, Section 6 concludes the paper.

\begin{figure}
    \centering
    \includegraphics[width=1.0\columnwidth]{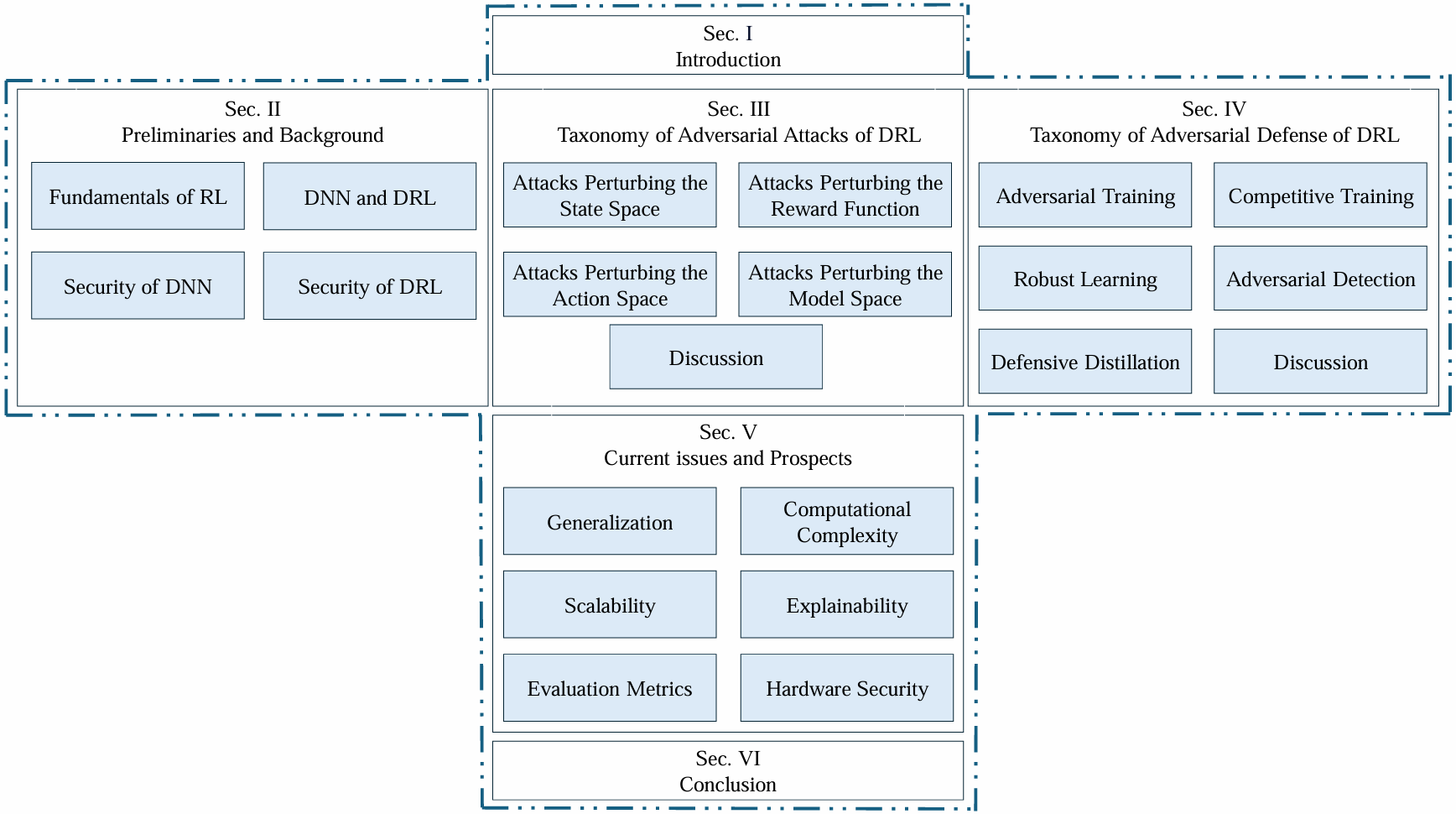}
                               Figure 1. Organization of this article   
\end{figure}

\label{introduction}
\section{Preliminaries and Background}
\subsection{Fundamentals of RL}
RL is a machine learning paradigm designed for sequential decision-making problems, aiming to enable an agent to learn policies that maximize cumulative rewards through interactions with the environment. Its core framework is typically formulated based on a Markov Decision Process (MDP), which characterizes the relationship between the agent’s decisions and the dynamic evolution of the environment through the definitions of states, actions, transition probabilities, and reward functions.
\subsubsection{Partially Observable Markov Decision Process}
A Partially Observable Markov Decision Process (POMDP) is an extended modeling framework for addressing the incomplete information problem in RL, describing the dynamic decision-making process of an agent under limited observations. Compared with fully observable Markov Decision Processes, POMDPs more closely reflect real-world learning scenarios, since in many practical tasks, an agent cannot directly access the complete state of the environment and must make decisions based on partial observations. A POMDP is typically represented by the septuple $X=(S, A, O, T, R, \Omega, Z)$, where:

 $S$:denotes the state space of the environment, representing all possible true states;
 
 $A$:denotes the action space, representing all possible actions available to the agent;
 
 $O$:denotes the observation space, representing the information received by the agent from the environment;
 
$T{:}S\times A \times S\to \left [ 0,1 \right ] $ is the state transition function, representing the probability of transitioning from state $s\in S$ to state $s^{\prime} \in S$ after taking action $a\in A$ ;

$R{:}S\times A\to \mathbb{R}$ is the reward function, specifying the immediate reward obtained by taking action $a$ in state $s$;

$\Omega{:}S\times A\times O\to[0,1]$ denotes the observation probability distribution, representing the probability of receiving observation $o \in O$ after taking action and arriving at state $s^{\prime}$;

 $Z$:represents the initial state distribution, defining the probability distribution over the environment’s initial states.
Within a POMDP framework, a single interaction with the environment can be represented by the quadruple $(s_{t},o_{t},a_{t},s_{t+1})$, where $s_{t}$ is the current true state of the environment,$o_{t}$ is the agent’s observation of that state, $a_{t}$ is the action taken by the agent, and $s_{t+1}$ is the next state following the transition. In this paper, although some prior studies model the environment as a fully observable Markov Decision Process, most works adopt POMDPs as a more general theoretical framework to address the incomplete information challenges faced by agents in realistic scenarios due to limited observations.
\subsubsection{Theory of Reinforcement Learning}
In the RL process, the objective of an agent is to learn a policy function $\pi(a|s)$, which defines the probability distribution over actions  given state . By optimizing $\pi$, the agent aims to identify the optimal policy $\pi^{*}$ that maximizes its expected cumulative return $G_{t}$, defined as:
\begin{equation}\pi^{*}=arg\max_{\pi}\mathbb{E}_{\tau\sim\pi}x[G_{t}]\end{equation}
\begin{equation}G_t=\sum_{t=0}^{|\tau|}\gamma^tR(s_t,a_t,s_{t+1})\end{equation}

where $R(s_t,a_t,s_{t+1})$ denotes the immediate reward obtained after the agent $a_{t}$ executes action $s_{t}$ in state and transitions to the next state . The trajectory $\tau =(s_0,a_0,r_0,s_1,a_1,r_1,...,s_{|\tau|})$is sampled from the distribution $\pi^X$ and represents the sequence of state-action-reward interactions between the agent and the environment, with $|\tau|$ denoting the trajectory length. The discount factor $\gamma\in[0,1)$ balances the relative importance of immediate and future rewards.

Within the theoretical RL framework, the state-value function $V^\pi$ quantifies the expected cumulative discounted return when starting from state $s \in S$ under a given policy $\pi$ , formally defined as:
\begin{equation}V^{\pi}(s)=\mathbb{E}_{\tau\sim\pi}x[G_{t}|s_{t}=s]|\end{equation}

To efficiently compute the state-value function, the Bellman equation is introduced, which recursively decomposes the cumulative return by expressing long-term objectives as a weighted sum of immediate rewards and future state values:
\begin{equation}V^{\pi}(s)=\mathbb{E}_{a\sim\pi,s^{^{\prime}}\sim T}\left[R(s,a)+\gamma V^{\pi}\left(s^{^{\prime}}\right)\right]\end{equation}

To further quantify the contribution of specific actions to cumulative returns, the action-value function $Q^\pi$ is defined, extending the state-value function to represent the expected cumulative discounted return of taking action $a \in A$ in state $s \in S$ and subsequently following policy $\pi$ :
\begin{equation}Q^{\pi}(s,a)=\mathbb{E}_{\tau\sim\pi}x[G_{t}|s_{t}=s,a_{t}=a]\end{equation}

which similarly satisfies the recursive Bellman equation:
\begin{equation}Q^{\pi}(s,a)=\mathbb{E}_{s^{^{\prime}}\sim T}\left[R(s,a)+\gamma\mathbb{E}_{a^{^{\prime}}\sim\pi}\left[Q^{\pi}\left(s^{^{\prime}},a^{^{\prime}}\right)\right]\right]\end{equation}

In a POMDP, the agent must make decisions based on limited observational information by maintaining a belief state that approximates the distribution over the true environment states. Specifically, the agent receives an initial state sampled from the initial state distribution $Z$ and selects an action $a \in A$ based on the current state; the environment then transitions to the next state $s^{\prime}$ according to the state transition function $T(s,a,s^{\prime})$ . Since the true environment state is not directly observable, the agent only receives an observation $o \in O$ according to the observation probability distribution $\Omega(s^{^{\prime}},a,o)$ and updates its state estimate using the previous state, current action, and new observation. By iteratively repeating this process, the agent gradually refines its estimation of the environment state and seeks to maximize long-term cumulative rewards through dynamic interactions. Figure 2 illustrates the dynamic interaction process between the agent and the environment in a POMDP.

\begin{figure}
    \centering
    \includegraphics[width=1.0\columnwidth]{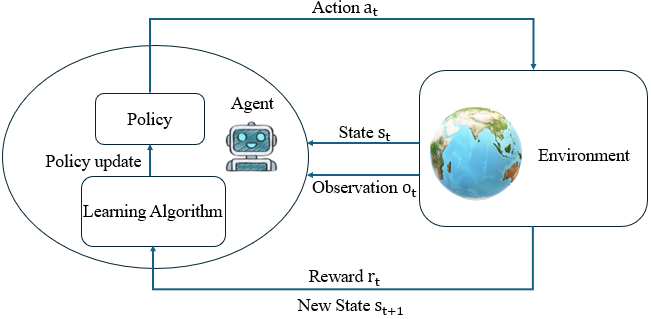}
    Figure 2. Basic process of the POMDP in RL
\end{figure}

\subsection{DNN and DRL}
DNNs have demonstrated powerful capabilities in representation learning for high-dimensional data\cite{ansuini2019intrinsic}. By integrating the feature extraction strengths of DNNs with the decision-optimization mechanisms of RL, agents can effectively handle complex high-dimensional state spaces and enhance their generalization capabilities in unknown environments, providing theoretical support for addressing real-world problems characterized by high complexity and partial observability.

\subsubsection{Deep Neural Network}
DNNs consist of multiple layers of interconnected neurons, enabling efficient feature learning through hierarchical processing and abstraction of data. The success of DNNs stems from their powerful nonlinear mapping capabilities, allowing end-to-end training via backpropagation without the need for manual feature engineering. Given an input data vector $x=[ \begin{array} {c}{x_{1},x_{2},...,x_{d}}\end{array}]\in\mathbb{R}^{d}$ , a DNN processes the data through successive layers of linear transformations followed by nonlinear activation functions. The output of the -th layer, $h_{l}$ , can be expressed as:
\begin{equation}\mathrm{h}_l=f_l(W_l\mathrm{h}_{l-1}+\mathrm{b}_l)\end{equation}
where $h_{l-1}$ denotes the output of the previous layer, $W_{l}$ is the weight matrix of the current layer, $b_{l}$ is the bias vector, and $f_{l}$ is the activation function. To efficiently train the parameters $\theta$ of a DNN, Stochastic Gradient Descent (SGD) is commonly employed to minimize a target loss function $L(\theta)$ . The parameter update rule is given by:
\begin{equation}\theta^{(t+1)}=\theta^{(t)}-\eta\nabla_{\theta}\mathcal{L}_{\mathcal{B}}\left(\theta^{(t)}\right)\end{equation}
where $\eta$ is the learning rate controlling the step size of each update, $B$ denotes a mini-batch of samples, and the gradient $\nabla_{\theta}L_{B}(\theta)$ is estimated based on the mini-batch $|B=\{(x_{i},y_{i})\}_{i=1}^{m}$ as:
\begin{equation}\nabla_{\theta}\mathcal{L}_{\mathcal{B}}=\frac{1}{m}\sum_{i=1}^{m}\nabla_{\theta}\mathcal{L}(x_{i},y_{i},\theta)\end{equation}
with  representing the number of samples in the mini-batch. By randomly selecting mini-batches, SGD significantly improves computational efficiency and, to some extent, mitigates the risk of being trapped in local minima.

\subsubsection{ Deep Reinforcement Learning}
DRL combines deep learning with RL, leveraging the powerful feature extraction capabilities of DNNs to enhance the performance of RL algorithms in high-dimensional state spaces and complex environments. In DRL, the environment state $s_{t}$ is mapped to input features, which are processed by a DNN to yield a policy $\pi(a|s_{t},\theta)$ , where $\theta$ denotes the parameters of the neural network. The agent optimizes these parameters to improve its performance in long-term tasks.

The core objective of RL is to update network parameters by minimizing a target function. In value-based DRL methods, the objective function is defined as:
\begin{equation}L(\theta)=\mathbb{E}_{(s_{t},a_{t},r_{t},s_{t+1})\sim\mathcal{D}}\left[\left(y_{t}-Q(s_{t},a_{t};\theta)\right)^{2}\right]\end{equation}

where $y_{t}$ denotes the target value and represents the experience replay buffer.

In contrast, policy-based DRL methods directly optimize the policy to maximize cumulative rewards, with the optimization objective defined as:
\begin{equation}J(\theta)=\mathbb{E}_{\tau\sim\pi_{\theta}}\left[\sum_{t=0}^{T}\gamma^{t}r_{t}\right]\end{equation}

Value-based DRL methods generally excel in discrete action spaces, whereas policy-based methods naturally accommodate continuous action spaces and often outperform in high-dimensional action domains. Recent state-of-the-art algorithms such as DSAC-T\cite{duan2023dsac}, CQL\cite{kumar2020conservative}, and IQL\cite{kostrikov2021offline} integrate the advantages of both value-based and policy-based approaches by constructing more stable objective functions and implementing efficient policy update mechanisms, thereby providing agents with robust and efficient decision-making capabilities in complex environments. Figure\ref{fig:DRL} illustrates the classification framework of mainstream DRL algorithms.

\begin{figure}
    \centering
    \includegraphics[width=1.0\columnwidth]{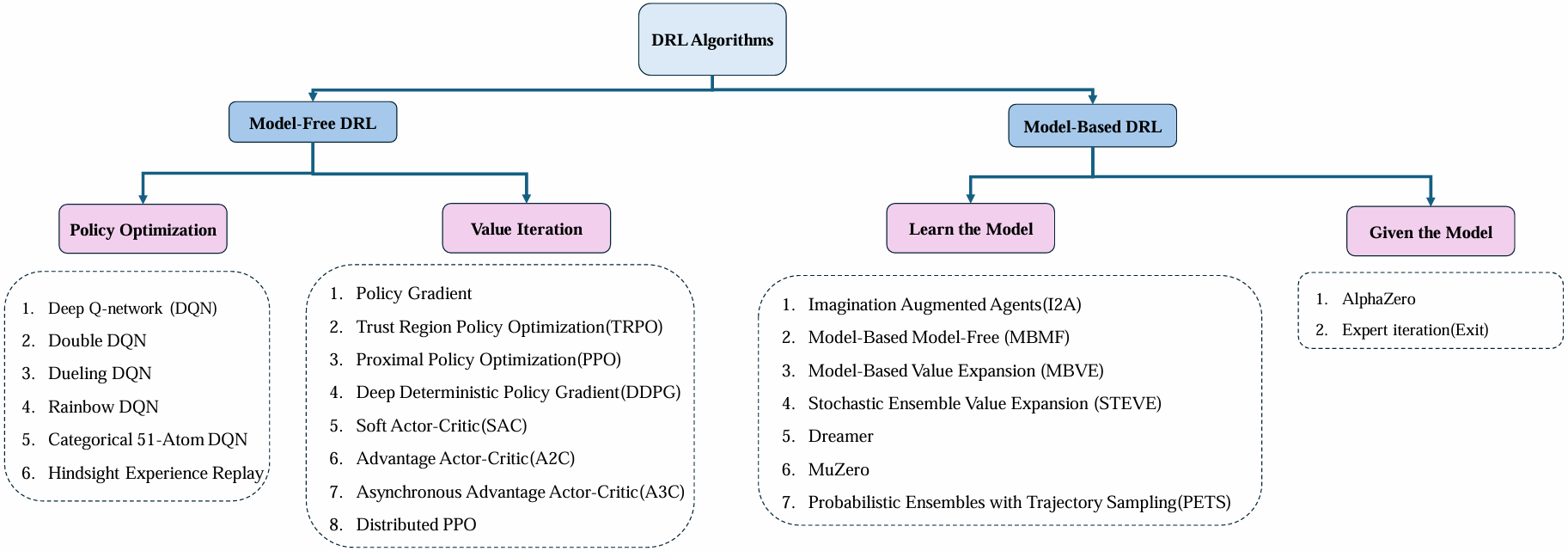}
    Figure 3. Taxonomy of Mainstream DRL Algorithms
\end{figure}

\subsection{Security of DNN}
With the widespread deployment and application of DNNs in critical domains, their potential security vulnerabilities have attracted significant attention from researchers. The high-dimensional nonlinear features and data-driven learning mechanisms of DNNs often make them unable to guarantee reliable inference when confronted with adversarial attacks or out-of-distribution data. Such intrinsic security flaws not only compromise the robustness and stability of DNN-based systems but may also lead to catastrophic consequences in high-risk decision-making scenarios, including autonomous driving, medical diagnostics, and financial risk management. The International Organization for Standardization (ISO) explicitly highlights in its recently released Artificial Intelligence Management System Standard (ISO/IEC 42001:2024) that in-depth investigation of DNN security challenges, adversarial machine learning theory, and defense strategy optimization holds significant theoretical value and practical relevance.

\subsubsection{Vulnerabilities of DNN}
The complex architecture of DNNs and their heavy reliance on large-scale training data expose them to numerous security vulnerabilities. These vulnerabilities not only arise from the intrinsic characteristics of the algorithms but are also closely related to the complexity of real-world environments and data properties. In terms of input perturbations, Goodfellow et al.\cite{goodfellow2014explaining} first observed that DNNs are highly sensitive to small perturbations in input data, a phenomenon that is particularly pronounced in high-dimensional spaces and is significantly amplified by adversarial examples. This vulnerability primarily stems from the strong dependence of DNNs on local gradients and the amplification effect of nonlinear features. Hendrycks et al.\cite{hendrycks2016baseline} further noted that DNNs heavily rely on the assumption of a consistent training data distribution, and any deviation in test data distribution can lead to substantial performance degradation. The black-box nature of DNNs also introduces non-negligible security risks. Papernot et al.\cite{papernot2017practical} demonstrated that, due to the lack of transparent decision logic, the behavior of DNNs under abnormal inputs is difficult to predict, posing potential threats to practical applications. Liu et al.\cite{liu2018survey} highlighted that the performance of DNNs is highly dependent on the quality and diversity of training data, making them particularly vulnerable to malicious data manipulation, such as data poisoning attacks. Any tampering with training data can lead to significant deviations in model decisions, thereby exacerbating security risks.

Zhao et al.\cite{zhao2024adversarial} systematically reviewed recent advances in adversarial training methods for defending against adversarial attacks. However, Rajhi et al.\cite{Rajhi2022_AdversarialTraining} pointed out that such methods typically require substantial computational resources, limiting their feasibility in resource-constrained environments. Moreover, Badjie et al.\cite{badjie2024adversarial} further demonstrated that even with state-of-the-art defense techniques, DNNs may still be susceptible to more sophisticated adversarial attacks.

\subsubsection{Adversarial Machine Learning}
Adversarial machine learning focuses on investigating the vulnerabilities of machine learning systems in adversarial environments and developing methods to enhance their robustness. Specifically, this field primarily addresses two core issues: first, identifying the potential weaknesses of existing models under adversarial scenarios; and second, constructing robust algorithmic frameworks to improve the models’ resistance to attacks. Adversarial attacks can occur at multiple stages of a model’s lifecycle. During the training phase, attackers may manipulate training data through data poisoning, thereby compromising the learning process. During the inference phase, adversarial examples can be crafted to bypass the model’s decision-making mechanism, leading to misclassification or other performance degradation.

In recent years, the scope of adversarial machine learning research has expanded beyond traditional attack and defense mechanisms to include model interpretability and generalization\cite{baniecki2024adversarial}. Researchers not only aim to delineate the capability boundaries of adversaries but also seek to comprehensively evaluate the specific impact of adversarial attacks on model performance and to develop theoretically grounded defense methods, thereby establishing a solid foundation for building secure and efficient artificial intelligence systems.

\subsubsection{Adversarial Attacks of DNN}
Adversarial attacks on DNNs involve applying subtle yet carefully crafted perturbations to normal samples, making it difficult for the model to distinguish between original and adversarial inputs, thereby producing erroneous predictions. Such attacks not only expose the security vulnerabilities of DNNs but also create opportunities for malicious exploitation. In recent years, numerous studies have revealed the susceptibility of DNNs in adversarial scenarios\cite{wei2024physical,shayegani2023survey}.

Based on the characteristics of the attack, adversarial attacks can be categorized in multiple ways. In terms of the target, they are generally classified as evasion attacks, poisoning attacks, extraction attacks, and backdoor attacks. Considering the attacker’s knowledge of the model, attacks can be categorized as white-box, gray-box, or black-box attacks. Additionally, depending on the stage at which the attack occurs, adversarial attacks can be further divided into training-phase attacks and inference-phase attacks. These attacks challenge the robustness boundaries of DNNs and have driven the development of new theoretical frameworks and algorithms aimed at enhancing model security. Nevertheless, current research still faces several challenges. On one hand, the lack of a unified evaluation framework makes it difficult to objectively compare the performance of different attack and defense methods, hindering the standardization process within the field\cite{croce2020reliable}. On the other hand, the effectiveness and robustness of adversarial examples in the physical world remain major research difficulties. For instance, environmental factors such as lighting conditions and viewpoint variations can weaken the effectiveness of adversarial examples\cite{athalye2018synthesizing}. Research on adversarial attacks has gradually driven the deep learning field toward greater security and robustness, laying a solid foundation for building trustworthy artificial intelligence systems. Figure 4 illustrates the fundamental classification framework for adversarial attacks on DNNs.

\begin{figure}
    \centering
    \includegraphics[width=1.0\columnwidth]{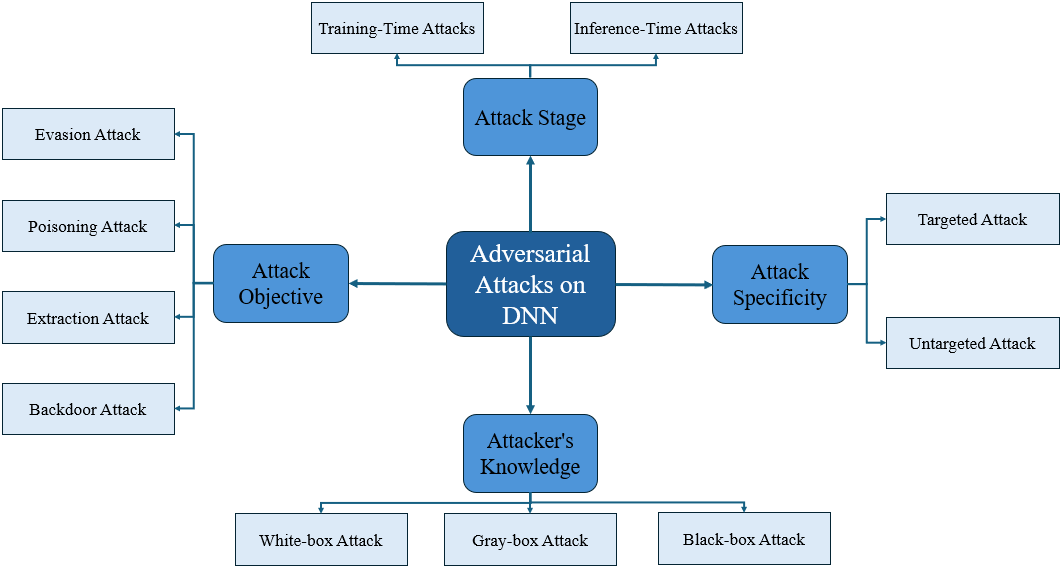}
Figure 4. Taxonomy of adversarial attacks on DNN
\end{figure}

\subsection{Security of DRL}
DRL optimizes decision-making policies through continuous interaction between agents and their environments. Compared with conventional deep learning methods, DRL not only relies on large-scale training data but also requires real-time adjustment of policies based on feedback in complex and dynamic environments, which introduces unique and intricate security challenges. Due to the high sensitivity of DRL’s training process and decision-making mechanisms to environmental feedback, malicious attackers may manipulate environmental states or reward signals to disrupt system behavior, potentially causing serious security risks. Therefore, systematically addressing these distinctive security challenges is of both theoretical significance and practical value for ensuring the robustness and reliability of DRL in real-world applications.

\subsubsection{Vulnerabilities of DRL}
The vulnerabilities of DRL primarily manifest as excessive sensitivity of agents to environmental states and reward signals, as well as adversarial risks during the training process. Studies have shown \cite{huang2017adversarial} that since DRL relies on environment-provided state information for decision-making, any manipulation of critical states or injection of false information by an attacker may lead the agent to misperceive the environment, causing policy failure. Experimental evidence indicates that in highly uncertain open environments, even minor state perturbations can result in significant performance degradation in well-trained DRL agents, with accuracy drops of up to 40–60\% \cite{zeng2022physics}. Manipulation of reward signals similarly poses a substantial security threat to DRL systems. As the direct driver of policy optimization, reward signals are highly sensitive and play a critical role in shaping agent behavior, particularly in continuous action spaces\cite{mo2022attacking}. When attackers carefully perturb the reward function, even small disturbances can cause the agent to learn policies that deviate significantly from intended objectives, potentially resulting in unpredictable and abnormal behaviors. Furthermore, since DRL models often rely on DNNs for feature extraction and decision mapping, the black-box nature and lack of interpretability of these models exacerbate the aforementioned risks, making it difficult for the system to effectively detect, localize, and mitigate potential security vulnerabilities.

\subsubsection{Robust RL}
Robust RL addresses the susceptibility of conventional RL to perturbations in uncertain environments by incorporating environment uncertainty modeling and worst-case optimization mechanisms, thereby enhancing the robustness and reliability of agents. Unlike traditional RL, which assumes environmental determinism and seeks to maximize expected returns, robust RL aims to optimize agent performance under worst-case scenarios, ensuring stability in adversarial or complex environments.

In the theoretical framework of robust RL, environmental uncertainty is typically modeled by introducing a perturbation set $\Delta$ , where $\delta\in\Delta$ represents potential disturbances affecting environment state transition probabilities or reward signals. Under this setting, the agent’s optimization objective can be formulated as a min–max problem\cite{morimoto2005robust}:
\begin{equation}\max_{\pi}\min_{\delta\in\Delta}J(\pi,\delta)\end{equation}

where $\pi$ denotes the agent’s policy, and $J(\pi,\delta)$ represents the expected return under policy $\pi$ and perturbation $\delta$ . The core idea of this optimization is that the agent selects an optimal policy $\pi$ that maximizes cumulative returns even under the worst-case perturbation.

Accordingly, the state-value function in robust RL is defined as:
\begin{equation}V^{\pi}(s)=\min_{\delta\in\Delta}\mathbb{E}_{a\sim\pi,s^{^{\prime}}\sim P(s^{^{\prime}}|s,a,\delta)}[R(s,a)+\gamma V^{\pi}(s^{^{\prime}})]\end{equation}

where $s$ is the current state, $a$ is the action, $s^\prime$ is the next state, $P(s^{^{\prime}}|s,a,\delta)$ denotes the state transition probability under perturbation $\delta$, $R(s,a)$ is the immediate reward, and $\gamma$ is the discount factor. This formulation reflects that the agent evaluates the worst-case return at each state.

The action-value function in robust RL further considers the impact of specific action choices on system robustness, defined as:
\begin{equation}Q^{\pi}(s,a)=\min_{\delta\in\Delta}\mathbb{E}_{s^{^{\prime}}\sim P\left(s^{^{\prime}}|s,a,\delta\right)}\left[R(s,a)+\gamma\max_{a}Q^{\pi}\left(s^{^{\prime}},a^{^{\prime}}\right)\right]\end{equation}

This definition captures how the selection of each action contributes to the agent’s performance under worst-case environmental perturbations.

\subsubsection{Adversarial Attacks of DRL}
Adversarial attacks, as techniques that induce erroneous model behaviors through subtle perturbations, have achieved significant success in supervised learning and have also been shown to pose severe threats to DRL agents\cite{behzadan2017vulnerability}. However, unlike single-step prediction tasks in supervised learning, RL aims to optimize long-term cumulative returns through interaction, meaning that the effects of adversarial attacks may require multi-step propagation to manifest. This temporal dependency not only increases the complexity of attack design and implementation but also complicates the evaluation of attack effectiveness. Within the POMDP framework, multiple core components—including observations\cite{standen2025adversarial}, states \cite{guo2024enhancing}, state transition functions\cite{liu2022false} , and reward functions\cite{zhang2021robust}—can serve as potential attack targets.

Given the high dependency of RL on environmental interaction, attackers must ensure that perturbations are both effective and stealthy, which imposes stringent requirements on attack design. Stealthiness requires avoiding conspicuous anomalies in the environment or agent behavior while accounting for the complexities of multi-agent interactions and dynamic environmental changes. Effectiveness demands that the perturbations continuously degrade the agent’s long-term cumulative returns in complex environments, thereby significantly impairing decision quality and policy performance. Based on these considerations, this study systematically explores the design of adversarial attacks within the RL framework, aiming to generate targeted perturbations against agents and improve system robustness under worst-case scenarios through training strategies, thus providing a critical foundation for enhancing DRL system resilience.
\label{PaB}
\section{Taxonomy of Adversarial Attacks of DRL}
Adversarial attacks on DRL can pose serious threats to the stability, performance, and reliability of various intelligent systems. To systematically understand these attack behaviors, this section aims to provide a comprehensive taxonomy of adversarial attacks in DRL. Based on the key functional components in the DRL process, attacks can be categorized into four main types. The majority of adversarial attacks primarily perturb the state space, aiming to manipulate the agent’s observations, the interactive environment, and the training data, thereby affecting the decision-making process. Relatively fewer attacks target the reward function or action space, intending to achieve their objectives by altering the agent’s learning goals or execution strategies. Additionally, model-targeted attacks disrupt the learning process fundamentally by modifying the model’s structure or parameters. Figure 5 illustrates the basic classification framework of adversarial attacks in DRL.

\begin{figure}
    \centering
    \includegraphics[width=1.0\columnwidth]{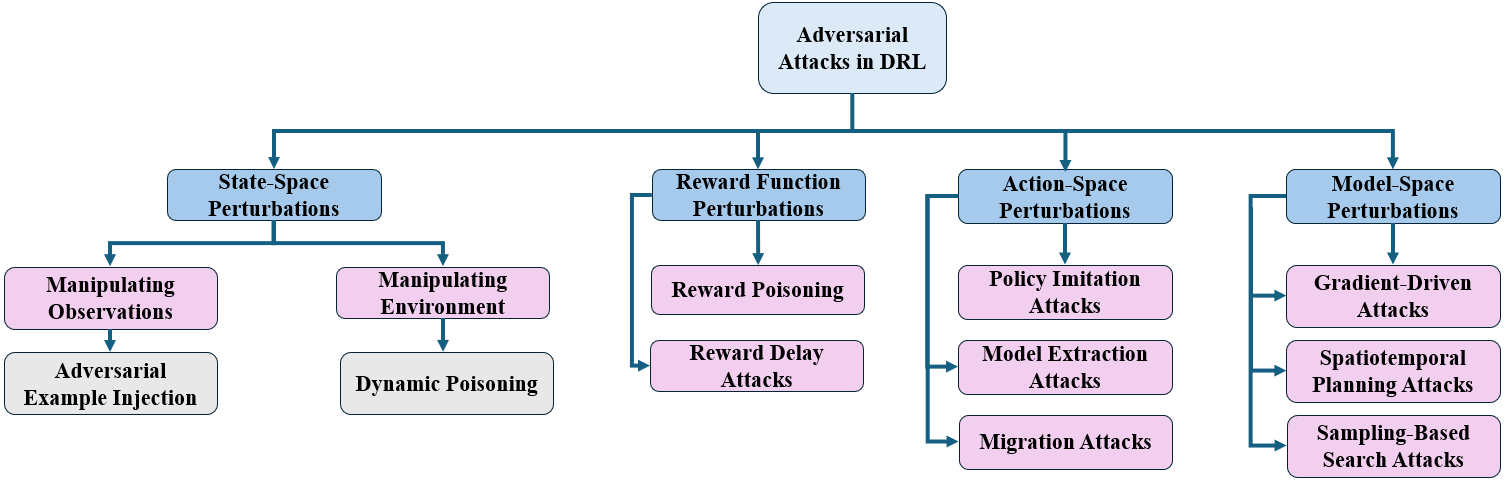}
   Figure 5.  Taxonomy of adversarial attacks on DRL
    \label{fig:ToAAoDRL}
\end{figure}

\subsection{Attacks Perturbing the State Space}
We further divide this subsection based on how the attacker accesses it.

\subsubsection{Manipulating the Observations}
Perturbations in the observation space represent one of the most extensively studied forms of adversarial attacks in DRL, with their theoretical foundation rooted in the vulnerabilities of DNNs observed in supervised learning scenarios\cite{goodfellow2014explaining}. Since DRL policy networks are essentially built upon DNNs, they are similarly sensitive to adversarial perturbations in perceptual inputs\cite{behzadan2017vulnerability}. The core objective of such attacks is to inject carefully crafted adversarial perturbations into the agent’s observations, causing the policy network to misinterpret state features and thereby induce suboptimal or even catastrophic decisions. The threat model assumes that the attacker can either directly access or indirectly infer the agent’s observation channels. To ensure stealthiness, the perturbations are generated under specific constraints. Typically, attackers employ optimization-based methods to create adversarial examples that minimize perceptibility, making them difficult to detect by human observers or monitoring systems while still effectively disrupting the agent’s decision-making process. The attack process generally involves two stages: the initialization stage and the exploitation stage. During initialization, the attacker trains a DRL model to develop an adversarial policy, guided by a crafted adversarial reward function, to generate disruptive behavior patterns. Subsequently, the attacker creates a replica of the target model and initializes its parameters. In the exploitation stage, adversarial inputs are generated to induce the target model to take specific actions according to the adversarial policy. Furthermore, in cross-model attack scenarios, the transferability of adversarial examples reflects the generalization of perturbations across heterogeneous DRL architectures. Fan et al.\cite{fan2022precise} demonstrated that adversarial attacks generated for a specific RL algorithm exhibit significant transfer effects across different algorithms. Experimental results indicate that attacks designed for DQN maintain substantial effectiveness when transferred to DQN, A2C, and PPO policies in strategy game environments. Similarly, attacks targeting A2C policies, though requiring increased attack frequency when transferred to other RL algorithms, can still markedly reduce the cumulative returns of the target algorithms. The cross-algorithm transferability of attacks reveals the pervasive vulnerabilities of various RL algorithms to adversarial perturbations, providing important theoretical insights for the robustness research of DRL systems. Figure 6 illustrates the attack process under the proposed framework.
\begin{figure}
    \centering
    \includegraphics[width=1.0\columnwidth]{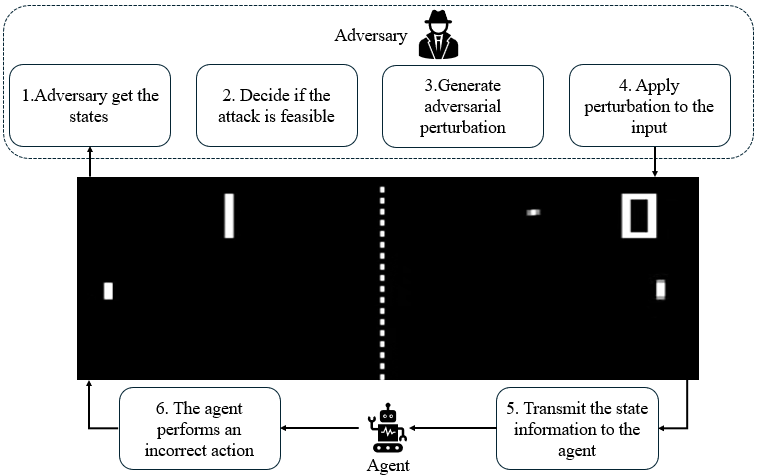}
   Figure 6. The adversary's attack process on the agent
    \label{fig:TAAPoTA}
\end{figure}

Zhang et al.\cite{zhang2020robust} proposed a robustness framework for state-observation adversarial perturbations in DRL by introducing the State-Adversarial Markov Decision Process (SA-MDP), systematically analyzing the properties of optimal policies under adversarially perturbed state observations. Their study revealed that under optimal adversarial perturbations, conventional deterministic policies may fail, and a globally optimal policy may not necessarily exist within the SA-MDP. You et al. \cite{qiaoben2024understanding} addressed the limitations of existing adversarial attack methods in terms of efficiency and effectiveness by proposing a unified framework based on functional space analysis. By categorizing optimization-based attacks into untargeted full-time attacks, strategically timed untargeted attacks, and targeted attacks, they theoretically demonstrated that targeted attacks possess stronger exploration capabilities in function space. They further noted that existing methods are constrained by the "pessimistic assumption"—i.e., assuming the performance of the attacked policy approximates that of the original policy—limiting their ability to effectively approach worst-case scenarios. To overcome this, the authors introduced a two-stage optimization method under the "optimistic assumption," dynamically generating adversarial perturbations to maximize cumulative reward loss by incorporating an intermediate deceptive policy and minimizing its KL divergence with the victim policy.

Traditional adversarial attack methods usually target the agent’s internal state. Hussenot et al. \cite{hussenot2017manipulating} proposed a more realistic attack paradigm that manipulates the observations provided by the environment rather than the internal state of the agent. They designed two novel untargeted attacks, namely frame-wise observation attacks and constant attacks, demonstrating the superiority and effectiveness of the latter in degrading agent performance. Additionally, a targeted attack was introduced, enabling precise manipulation of observations to compel the agent to execute a specific policy. Experiments in the Space Invaders environment showed that through frame-wise targeted attacks, the performance of DQN could be substantially improved, approaching the average performance of Rainbow\cite{hessel2018rainbow}.

Most existing RL adversarial attack methods rely on environment-specific rules or white-box access, requiring attackers to obtain the victim agent’s policy information, which poses practical limitations. Yamabe et al.\cite{yamabe2024behavior} introduced a novel State-Adversarial Imitation Attack (SAIA) that manipulates agent behavior in black-box and policy-agnostic settings, without requiring access to the victim’s policy or environment-specific heuristics. SAIA models the attack as a bi-level optimization problem and combines inverse RL with RL techniques to effectively mimic target policies and manipulate behaviors. Experimental results in Meta-World robotic manipulation tasks and autonomous driving environments demonstrated that SAIA outperforms baseline methods under black-box settings, achieving higher attack success rates across most tasks. Even without policy information, the method can effectively control agent behavior, whereas conventional random perturbation methods fail to achieve comparable results.

Most RL methods assume accurate state observations during training, which often does not hold in real-world scenarios. Although such methods can enhance algorithmic robustness, they remain susceptible to noise. In contrast, Distributional RL models the full return distribution rather than just its expectation, providing a more accurate representation of environmental uncertainty and improving robustness in noisy settings. Sun et al.\cite{sun2020exploring} constructed the State-Noisy Markov Decision Process (SN-MDP) and demonstrated the contraction and convergence properties of the distributional Bellman operator under both stochastic and adversarial noise. Gradient norm analysis further showed that the distributional RL loss, parameterized via categorical distributions and KL divergence, maintains a stable gradient upper bound during training, preventing potential gradient explosion.

Existing methods primarily guide agents to select suboptimal actions at each step but may not effectively reduce cumulative rewards. To address this, Chan et al.\cite{chan2020adversarial} introduced the Static Reward Impact Map (SRIM) to quantify the contribution of each feature to cumulative rewards, providing guidance for constructing adversarial samples. Specifically, SRIM estimates the impact of features on rewards and selects those with the greatest influence for perturbation. Compared to one-shot and random attacks, the proposed method significantly reduces the cumulative rewards of target agents under both white-box and black-box settings. Moreover, SRIM can consistently evaluate feature importance and maintain attack effectiveness across different agents, allowing attackers to launch effective attacks even without full knowledge of the target agent. The method also considers feature correlations during attacks, overcoming the limitations of conventional methods that perturb each feature independently.

In cooperative multi-agent RL (C-MARL) scenarios, agents have access only to local observations and must collaborate to achieve shared objectives. Limited observability and inter-agent dependencies pose a core challenge for effective cooperation under partial information. A key research paradigm in C-MARL is Centralized Training with Decentralized Execution (CTDE), where agents share global information during training for centralized policy optimization but act independently based on local observations during execution. Despite coordination during training, agents often rely solely on local information at execution time.

Addressing the security of C-MARL, Chen et al.\cite{chen2022marnet} proposed a novel attack method, MARNet, specifically targeting C-MARL systems. MARNet introduces specialized triggers, action poisoning, and reward manipulation, allowing agents to perform normally in clean environments but significantly degrading performance under malicious conditions. The method consists of three primary modules—trigger design, action poisoning, and reward manipulation. Triggers are carefully embedded in the environment, ensuring that even if only a few agents can observe them, these agents are effectively induced to execute worst-case actions.

\subsubsection{Manipulating the Environment}
In adversarial attacks on RL, environment manipulation refers to scenarios where the attacker indirectly influences an agent’s behavior by modifying certain environmental features or dynamics. The key idea behind this attack is to guide the agent toward decisions favorable to the attacker through adjustments in the environment. Common approaches include: altering the reward function to grant undue high rewards or penalties under specific circumstances; modifying state transition rules, causing the agent to enter unintended states; introducing random noise or perturbations that disrupt the agent’s observations and decision-making process; and embedding “triggers” or external signals that influence policy selection. Through these means, the attacker does not directly interfere with the agent’s actions but gradually induces incorrect choices via environmental changes, thereby undermining system performance and stability.

Xu et al.\cite{xu2021transferable} proposed a Transferable Environment-dynamics Poisoning Attack (TEPA) framework, extending the traditional training-time attack dimension from reward signal manipulation to covert control over environment state transitions. The framework achieves closed-loop optimization of attack strategies via a bi-level Markov Decision Process. TEPA quantifies the stealth cost of environmental dynamics modification using KL divergence and employs DRL algorithms to generate optimal attack strategies under minimal perturbations. Unlike existing environment poisoning methods that are limited to specific algorithms, TEPA demonstrates the feasibility of cross-algorithm attacks by training attack strategies under white-box settings and transferring them to black-box victim agents, highlighting the role of exploration intensity in the transferability of attack strategies.

Differing from traditional environment poisoning attacks that modify only the reward function, Rakhsha et al.\cite{rakhsha2020policy} employed an optimization framework to minimize attack costs and provided theoretical analysis for independent poisoning of rewards and state transitions. This study expanded the scope of attacks by proposing a joint manipulation strategy targeting both rewards and state transitions. Experimental results show that in offline settings, attackers can influence agents’ planning by modifying the environment, while in online settings, attackers can directly alter feedback signals to force agents to execute target policies. The findings indicate that attackers can successfully coerce target agents into executing desired strategies at relatively low cost, with notable effectiveness across RL algorithms using discounted or average reward criteria.

Existing single-agent environment poisoning attacks mainly focus on manipulating reward signals or observations to influence agent behavior. However, these approaches face significant limitations in multi-agent environments, particularly when the internal policies of individual agents are unavailable, making it difficult to achieve desired outcomes. To address this, Bector et al.\cite{bector2021poisoning} proposed an innovative Collective Environment Poisoning (CEP) method, specifically designed for training-time attacks in multi-agent systems. CEP can simultaneously attack a group of learning agents and guide the entire cohort toward a common target behavior while minimizing environmental changes. By incrementally adjusting and parameterizing the environment, CEP enables attacks to transfer across groups of varying sizes without relying on the internal structure of any single agent. The method effectively directs the agent cohort to perform attacker-specified behaviors without significantly altering environmental dynamics. Compared to traditional behavior-linkage attacks, CEP demonstrates strong transferability across diverse group sizes and achieves higher attack success rates with lower environmental modification costs.

In conventional RL, agents learn optimal policies through interactions with the environment, a process that typically requires extensive interaction data. Due to the complexity of data collection and privacy concerns, RL faces substantial challenges. To address this, Federated Learning (FL) has been proposed as a framework that allows multiple agents to collaboratively train models while preserving data privacy. Despite its advantages in privacy protection and communication cost reduction, the decentralized nature of FL introduces significant security vulnerabilities. Ma et al.\cite{ma2023local} proposed a local environment poisoning attack targeting federated RL systems. The method manipulates the local environment observations of malicious agents to produce states, actions, or rewards harmful to global training, coercing the server into learning a compromised global model. This attack considers the complexity of the target environment and can disrupt the global model using only local observations, without requiring full knowledge of the environment’s Markov Decision Process. Experiments across different policy gradient algorithms show that a small number of malicious agents can effectively poison the entire federated learning system, achieving higher attack success rates compared to baseline attack methods.

\subsection{Attacks Perturbing the Reward Function}
Adversarial perturbation attacks targeting the reward function have emerged as a significant threat to safety-critical RL systems. These attacks distort the policy gradient update direction by introducing inducive bias into the reward signal while keeping the environment dynamics unchanged. Unlike state observation perturbations or action-space interference, reward function attacks exhibit higher stealth and strategic disruption: attackers may directly manipulate the reward generation mechanism in a white-box setting or exploit vulnerabilities in policy gradient estimation to induce indirect perturbations in a black-box setting. Fundamentally, the attack deconstructs the mathematical expectation structure of the RL objective function, designing perturbation constraints such that the optimal policy converges to a pre-specified suboptimal policy under the perturbed reward function. Studies show that such attacks can induce systematic failures in DRL agents in high-risk scenarios, including policy collapse and reward hijacking, with attack effects propagating exponentially with the number of training episodes.

Addressing reward poisoning in offline multi-agent RL systems, Wu et al.\cite{wu2023reward} proposed a novel attack framework for cooperative multi-agent learning scenarios. Unlike single-agent settings, which require independently attacking each agent, the attacker modifies the reward signals in the offline dataset to enforce a Markov Perfect Dominant Strategy Equilibrium (MPDSE), ensuring that all rational agents inevitably follow the target policy. Theoretical analysis demonstrates that this method significantly reduces the complexity of multi-agent cooperative attacks by constructing a backward induction optimization model and efficiently solving the minimal-cost attack using linear programming.

In the domain of large models, research on systemic vulnerabilities under the RLHF framework has revealed multidimensional threat paths from reward model poisoning attacks. To address this risk, Wang et al.\cite{wang2024rlhfpoison} proposed RankPoison, a reward model poisoning attack targeting RLHF mechanisms. The method first selects adversarial samples exhibiting preference reversals to identify conflicts between human labels and attack objectives, then dynamically evaluates the adversarial impact of each sample based on reward model loss differences to evade detection, and finally amplifies perturbations in a directed manner by maximizing measurable statistical deviations, such as text length.

Although existing reward poisoning attacks can manipulate models via preference reversal, their attack paths remain constrained by explicit data contamination dependencies and the detectability of specific semantic trigger patterns. To overcome this limitation, Sarkar et al.\cite{sarkar2020reward} first identified the vulnerability of DRL algorithms to delayed reward attacks and proposed an Asynchronous Reward Injection (ARI) framework to break the temporal synchronization assumption. The authors constructed a dual-mode threat model of non-targeted and targeted attacks: the former degrades DQN agent performance to near-random levels in Atari benchmarks via random reward delays or selective omission, while the latter designs proxy reward functions to successfully induce target actions in 75\% of target states. Experiments indicate that even with secure timestamp verification defense mechanisms, attackers can substantially reduce learning efficiency in adversarial environments through temporally consistent reward shift attacks.

Ma et al.\cite{ma2024reward} proposed ReLara, a novel dual-agent cooperative reward shaping framework that addresses RL efficiency bottlenecks in sparse and delayed reward scenarios. ReLara models the reward function as a partially observable Markov decision process, where a reward agent generates forward-looking dense reward signals based on state-action pairs, and the policy agent optimizes strategies through a weighted combination of environment and shaping rewards. The framework employs a double-Q network architecture and independent experience replay pools to ensure decoupled optimization between reward prediction and policy update. Theoretically, ReLara demonstrates an adaptive balancing mechanism: early-stage random reward injections enhance exploration, while later-stage information-gain rewards enable directed policy optimization, mitigating convergence oscillations caused by reward sparsity in traditional methods.

The aforementioned methods do not consider direct manipulation of reward values in offline static datasets, limiting attack efficiency. To address this, Xu et al.\cite{xu2024universal} proposed a general black-box reward poisoning attack for deep offline RL, termed Policy Contrast Attack (PCA). This attack reconstructs the reward function via an adversarial reward engineering framework, making high-performing strategies in the dataset appear inefficient under adversarial conditions, while enhancing pseudo-reward representations of low-performing strategies. This misleads pessimistic optimal learning algorithms into outputting suboptimal policies, addressing limitations of existing methods that rely on perturbation priors, apply only to discrete rewards, or cannot handle optimal policy shifts. Figure 7 illustrates the framework structure of reward poisoning attacks.
\begin{figure}
    \centering
    \includegraphics[width=1.0\columnwidth]{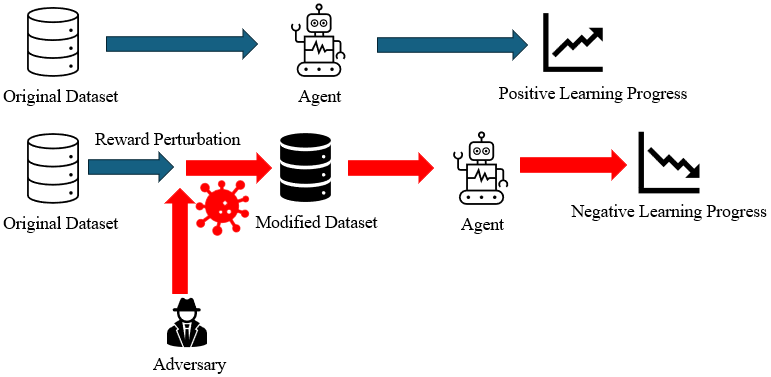}
   Figure 7. Reward poisoning attack framework
    \label{fig:RPAF}
\end{figure}

\subsection{Attacks Perturbing the Action Space}
The action space serves as the core interface through which agents interact with the environment to optimize policy decisions. Adversarial attacks targeting the action space operate by directly perturbing the agent’s action outputs or injecting deceptive signals into the action selection mechanism, inducing suboptimal or catastrophic behaviors. Compared with state-space attacks that manipulate environmental observations, action-space attacks directly exploit vulnerabilities within the agent’s decision-making pipeline and can often bypass perception-based defense mechanisms. These attacks exhibit significant stealth in continuous control tasks such as robotic manipulation and autonomous navigation, where minor deviations in action trajectories can accumulate over time to trigger systemic failures.

Lee et al.\cite{lee2020spatiotemporally} proposed a spatiotemporally constrained action-space adversarial attack framework, marking the first extension of DRL attack dimensions from traditional state-space perturbations to covert action-space manipulation. The framework introduces Myopic Action Space (MAS) and Look-ahead Action Space (LAS) attacks to enhance attack efficacy. MAS minimizes immediate rewards within single-step action spaces via greedy projected gradient descent, whereas LAS incorporates temporal coupling constraints and dynamic programming, combining surrogate environment simulation with mixed-norm projection to allocate the attack budget non-uniformly across critical time steps and action dimensions.

Luo et al.\cite{luo2024provably} addressed action-space adversarial attacks in continuous RL agents by proposing LcBT, an efficient black-box attack framework that integrates Monte Carlo Tree Search with lower confidence bound techniques to tackle the complexity of searching for attack strategies in continuous action spaces. This study first defines the target action space and policy set within the threat model, discretizes the continuous action space into searchable node structures via a hierarchical action coverage tree, and optimizes perturbation selection using trajectory-driven Q-value estimation and importance sampling. Theoretical analysis demonstrates that when the agent’s dynamic regret satisfies sublinear conditions, LcBT can force convergence to the target policy with sublinear attack cost.

Conventional adversarial attack methods face difficulties dynamically selecting key agents and effectively disrupting team coordination strategies in continuous action spaces. When the set of attacked agents changes, existing frameworks require retraining to adapt to new targets. To overcome this limitation, Zhou et al.\cite{zhou2024adversarial} proposed AMCA, an adversarial attack framework for MARL models, which extends the attack dimension from fixed victim agents to dynamically selected key agents. AMCA identifies critical agents and generates worst-case joint actions through gradient information and differential evolution. Unlike attacks focusing on individual agent perturbations, AMCA dynamically selects key agents and generates targeted adversarial observations, significantly undermining team coordination strategies. Experimental results indicate that AMCA outperforms existing methods in perturbation capability, achieves attacks on dynamic victim agents without pretraining, and demonstrates stronger efficacy in industrial-relevant environments.

\subsection{Attacks Perturbing the Model Space}
In adversarial research on RL systems, the model space attack refers to scenarios where an attacker interferes with the decision-making process by altering the model’s structure, parameters, or weights. Unlike attacks that directly manipulate model inputs or training data, model space attacks affect the internal architecture or optimization process, coercing the model to produce suboptimal or biased outputs even when presented with normal inputs. Techniques for attacking the model space include, but are not limited to, introducing small perturbations to model weights, modifying the training procedure, or inserting malicious features into intermediate layers to influence final decisions. Compared with input data poisoning or environment poisoning attacks, model space attacks are more covert and robust, as they act directly on the core of the model; even under standard inputs, the model can be forced to exhibit unintended behaviors. A key characteristic of model space attacks is their high specificity and flexibility: by precisely manipulating layer-wise weights, attackers can induce prediction errors or performance degradation without significantly altering the input, which is particularly pronounced in DNNs.

Behzadan et al.\cite{behzadan2019adversarial} investigated the transferability of adversarial examples in imitation-based DRL policies and proposed an adversarial attack method targeting DRL policy confidentiality. This method iteratively queries the original model and uses Deep Q-Learning from Demonstrations (DQfD) to mimic the target policy, generating an approximate policy from which adversarial examples are crafted and then transferred to the original policy. Experiments using FGSM for perturbation generation demonstrated significant transferability of adversarial examples between the mimic and original policies, with transfer success positively correlated with the amount of demonstration data. This supports the hypothesis that the decision boundaries of the mimic and original policies are similar. Moreover, a higher volume of demonstration data reduces distributional discrepancies between policies, enhancing the efficiency of adversarial example transfer and providing a feasible path for black-box attacks without direct access to model parameters.

While significant progress has been made in extraction attacks against supervised deep learning models, model extraction attacks targeting DRL remain limited. Traditional methods often rely on model confidence outputs or gradient information; however, deployed DRL models typically expose only discrete actions, providing insufficient information for effective extraction. Furthermore, supervised model extraction techniques struggle to capture the temporal decision-making characteristics of DRL, as independent input samples cannot fully reflect the dynamic dependencies of a Markov Decision Process (MDP). To address these limitations, Zhuang et al.\cite{zhuang2024stealthy} proposed Stealthy Imitation, a model space perturbation attack that achieves functional policy extraction via black-box queries. Under constraints of limited environment interaction and no prior knowledge of input ranges, the attacker constructs a state-space estimation model based on diagonal Gaussian distributions and optimizes the query distribution through an iterative reward-guided mechanism. The method replaces policy training with behavioral cloning and introduces a dual-reward model based on adversarial training to dynamically evaluate discrepancies between the victim and attack policies, guiding the query distribution to converge to the victim policy’s true state space. Chen et al.\cite{chen2021stealing} introduced a novel extraction attack against black-box DRL models. The approach theoretically demonstrates the equivalence between model extraction and imitation learning and establishes a two-stage attack framework. First, a RNN classifier is trained on temporal action sequence features to accurately identify the training algorithm of the target model. Then, using a Generative Adversarial Imitation Learning (GAIL) framework, a surrogate model is reconstructed from the target model’s environment interaction trajectories through adversarial training between the policy and discriminator networks, achieving equivalent decision-making capability and high behavioral fidelity. Experiments indicate that this method can precisely replicate a highly similar DRL model by observing external behavior alone, without access to internal structures, providing new insights for model protection, adversarial attack design, and privacy preservation.

\subsection{Discussion}
This section systematically analyzes adversarial attack paradigms targeting deep DRL and provides a detailed comparison and summary of different attack methods, as shown in Table 1. Based on the attacker’s access to different components, attacks can be categorized into four types: state space attacks, action space attacks, reward function attacks, and model space attacks. State space attacks perturb the environment’s state or the agent’s observation inputs, thereby interfering with the agent’s perception of the environment. Action space attacks directly manipulate the agent’s execution mechanism or action selection process, disrupting its behavioral decisions. Reward function attacks mislead the agent’s policy optimization by altering reward signals or preference feedback, steering it toward suboptimal or erroneous strategies. Model space attacks intrude into the core parameters or architecture of DRL models, modifying their intrinsic knowledge or learning mechanisms and thereby affecting the agent’s learning process and decision-making performance.

In practical scenarios, attacks on DRL systems not only require high stealth but also must meet constraints on low query cost. Future research should focus on developing specialized and efficient attack frameworks for DRL, enabling a more comprehensive assessment of DRL system robustness while balancing attack efficacy and resource consumption.

\begin{table*}[H]
  \centering
   Table 1. Summary of Adversarial Attacks methods on DRL
    \resizebox{\linewidth}{!}{
    \begin{tabular}{|c|c|m{5.25cm}<\centering|c|c|c|c|c|c|c|c|} 
    \hline
    \multirow{2}{*}{\textbf{Paper}} & \multirow{2}{*}{\textbf{DRL Technique}} & \multirow{2}{*}{\textbf{Environment}} & \multicolumn{2}{c|}{\textbf{Test-time Attacks}} & \multicolumn{2}{c|}{\textbf{Train-time Attacks}} & \multicolumn{4}{c|}{\textbf{Attack Objective}}  \\
    \cline{4-11}
          &       &       & White box & Black box & White box & Black box & State & Action & Reward & Model \\
    \hline
    Behzadan et al. {[}37{]} & DQN   & Atari (Pong) & $\checkmark$     & $\checkmark$     & $\times$     & $\checkmark$     & $\checkmark$     & $\times$     & $\times$     & $\times$\\
    \hline
    Fan et al. {[}42{]} & DQN, A2C & Atari (Pong, Breakout) & $\checkmark$     & $\checkmark$     & $\times$     & $\times$     & $\checkmark$     & $\times$     & $\times$     & $\times$ \\
    \hline
    Zhang et al. {[}43{]} & DQN, PPO, DDPG & \makecell{Atari (Pong, Freeway, etc.),\\ MuJoCo (Hopper, Walker2d, etc.)} & $\checkmark$     & $\times$     & $\checkmark$     & $\times$     & $\checkmark$     & $\times$     & $\times$     & $\times$ \\
    \hline
    You et al. {[}44{]} & DQN, A2C, PPO & \makecell{Atari (Pong, Qbert, etc.),\\ MuJoCo (Ant, Hopper, etc.)} & $\checkmark$     & $\times$     & $\times$     & $\times$     & $\checkmark$     & $\times$     & $\times$     & $\times$ \\
    \hline
    Hussenot et al. {[}45{]} & DQN, Rainbow & Atari (Space Invaders) & $\checkmark$     & $\times$     & $\times$     & $\times$     & $\checkmark$     & $\times$     & $\times$     & $\times$ \\
    \hline
    Yamabe et al. {[}47{]} & SAC   & Meta-World, highway-env & $\times$     & $\checkmark$     & $\times$     & $\times$     & $\checkmark$     & $\times$     & $\times$     & $\times$ \\
    \hline
    Sun et al. {[}48{]} & \makecell{C51, QRDQN,\\ DSAC, D4PG} & \makecell{MuJoCo (CartPole, MountainCar, etc.),\\ Atari (Breakout, Qbert)} & $\times$     & $\times$     & $\checkmark$     & $\times$     & $\times$     & $\times$     & $\times$     & $\checkmark$ \\
    \hline
    Chan et al. {[}49{]} & DQN   & Atari (Freeway, Pong, etc.) & $\checkmark$     & $\checkmark$     & $\times$     & $\times$     & $\checkmark$     & $\times$     & $\times$     & $\times$ \\
    \hline
    Chan et al. {[}50{]} & \makecell{VDN, QMIX,\\ COMA, SMAC} & Predator Prey, StarCraft & $\checkmark$     & $\checkmark$     & $\checkmark$     & $\times$     & $\checkmark$     & $\checkmark$     & $\checkmark$     & $\times$ \\
    \hline
    Xu et al. {[}51{]} & \makecell{Q-learning, DQN,\\ TD3, Sarsa} & 3D Grid World, 2-Goal Grid World & $\checkmark$     & $\checkmark$    & $\checkmark$     & $\checkmark$     & $\checkmark$     & $\times$     & $\times$     & $\times$ \\
    \hline
    Rakhsha et al. {[}52{]} & \makecell{Q-learning,\\ DQN, Sarsa} & Grid World & $\checkmark$     & $\checkmark$     & $\checkmark$     & $\checkmark$     & $\checkmark$     & $\times$     & $\checkmark$     & $\times$ \\
    \hline
    Becror et al. {[}53{]} & General RL & 3D Grid World & $\checkmark$     & $\checkmark$    & $\times$     & $\checkmark$    & $\checkmark$     & $\times$     & $\times$     & $\times$ \\
    \hline
    Ma et al. {[}54{]} & PG, PPO & \makecell{Gym (CartPole,\\ Inverted Pendulum, etc.)} & $\checkmark$     & $\checkmark$     & $\checkmark$     & $\checkmark$     & $\checkmark$     & $\times$     & $\checkmark$     & $\times$ \\
    \hline
    Wu et al. {[}55{]} & Offline MARL & Markov Games & $\times$     & $\times$     & $\times$     & $\checkmark$    & $\times$     & $\times$     & $\checkmark$     & $\times$ \\
    \hline
    Wang et al. {[}56{]} & RLHF  & Text Generation Tasks & $\times$     & $\times$     & $\times$     & $\checkmark$     & $\times$     & $\times$     & $\checkmark$     & $\times$ \\
    \hline
    Sarkar et al. {[}57{]} & DQN   & Atari (Pong, Breakout) & $\times$     & $\times$     & $\times$     & $\checkmark$     & $\times$     & $\times$     & $\checkmark$     & $\times$ \\
    \hline
    Ma et al. {[}58{]} & ReLara & \makecell{MuJoCo, Arm Robot,\\ Physical Control} & $\times$     & $\times$     & $\times$     & $\checkmark$   & $\times$     & $\times$     & $\checkmark$     & $\times$ \\
    \hline
    Xu et al. {[}59{]} & Offline RL & D4RL (HalfCheetah, Hopper, etc.) & $\times$     & $\checkmark$     & $\times$     & $\checkmark$     & $\times$     & $\times$     & $\checkmark$     & $\times$ \\
    \hline
    Lee et al. {[}60{]} & DQN, PPO & Gym (Lunar Lander, Bipedal Walker) & $\times$     & $\times$     & $\times$     & $\checkmark$     & $\times$     & $\checkmark$     & $\times$     & $\times$ \\
    \hline
    Liao et al. {[}61{]} & \makecell{DDPG, PPO,\\ TD3} & \makecell{slider on a rail vehicle,\\ vehicle with complex control} & $\checkmark$     & $\checkmark$     & $\checkmark$     & $\checkmark$     & $\times$     & $\checkmark$     & $\times$     & $\times$ \\
    \hline
    Zhou et al. {[}62{]} & \makecell{MADDPG, FACMAC,\\ MAMuJoCo, MPE} & CMUE, MAMuJoCo, MPE & $\checkmark$     & $\checkmark$     & $\checkmark$     & $\checkmark$    & $\checkmark$     & $\checkmark$     & $\times$     & $\times$ \\
    \hline
    Behzadan et al. {[}63{]} & \makecell{DQN, A2C,\\ PPO} & Cart-Pole & $\checkmark$     & $\checkmark$     & $\checkmark$     & $\checkmark$    & $\times$     & $\times$     & $\times$     & $\checkmark$ \\
    \hline
    Zhuang et al. {[}64{]} & \makecell{DQN, PPO,\\ A2C} & \makecell{MuJoCo (Hopper,\\ Walker2d, HalfCheetah)} & $\times$     & $\checkmark$     & $\times$     & $\checkmark$    & $\checkmark$     & $\checkmark$     & $\times$     & $\checkmark$ \\
    \hline
    Chao et al. {[}65{]} & \makecell{DQN, PPO,\\ ACER, ACKTR} & Gym (CartPole), Atari (Pong) & $\times$     & $\checkmark$     & $\times$     & $\checkmark$     & $\times$     & $\checkmark$    & $\times$     & $\checkmark$\\
    \hline
    \end{tabular}}
  \label{tab:drl_attack}
\end{table*}

\label{ToAAoD}
\section{Taxonomy of Adversarial Defense of DRL}
Conventional RL methods struggle to cope with adversarial perturbations in complex environments. As a result, adversarial defense and robustness have become critical research topics in the field of RL security in recent years. In this section, we construct a classification system based on the intervention level and theoretical framework of defense strategies, systematically reviewing five core categories: adversarial training, competitive training, robust learning, adversarial detection, and defensive distillation. Each category of defense strategy is analyzed in the context of specific threat scenarios and operational constraints, with discussions on their theoretical boundaries and implementation paradigms. This provides multidimensional theoretical support and practical guidance for enhancing the trustworthiness and robustness of RL. Figure 8 presents a comprehensive classification framework of DRL defense strategies.

\begin{figure}
    \centering
    \includegraphics[width=1.0\columnwidth]{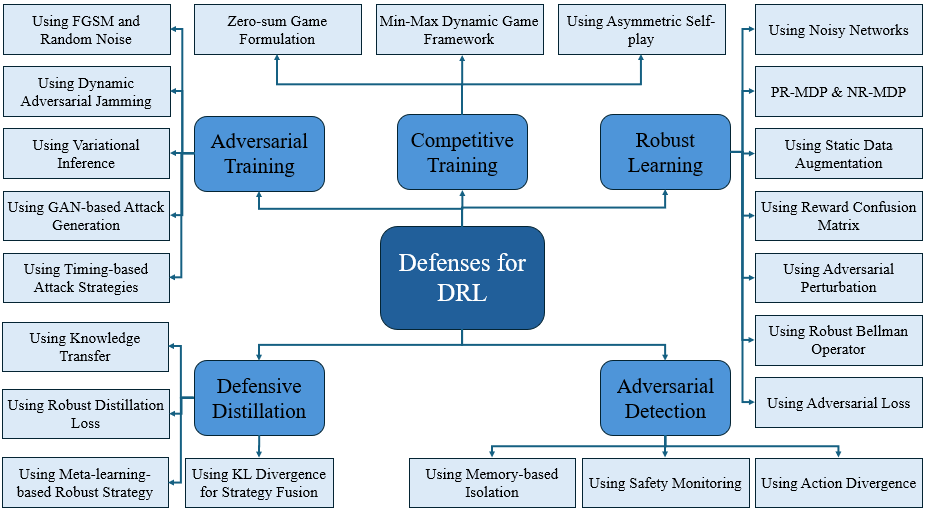}
   Figure 8. Taxonomy of Adversarial Defenses for DRL
    \label{fig:ToADfDRL}
\end{figure}

\subsection{Adversarial Training}
Adversarial training is a widely used defense mechanism that enhances model robustness by incorporating adversarial perturbations during the learning process. In RL, adversarial training generates adversarial samples and integrates them into the training phase, enabling agents to adapt to adversarial environments and maintain strong performance even under perturbations, thereby improving resilience against unknown attacks. Kos et al.\cite{kos2017delving} introduced an adversarial retraining paradigm for DRL, which constructs hybrid training samples by combining targeted perturbations generated via the FGSM method with random noise, and performs dynamic adversarial training during policy optimization. Agents trained under this framework not only mitigate performance degradation against homogeneous adversarial attacks but also demonstrate stable resistance to perturbations of varying intensity. Traditional adversarial training typically relies on online sample generation, which may lead to uncontrollable behaviors in safety-critical scenarios. To address this limitation, Liu et al.\cite{liu2023towards} proposed the SAFER algorithm, which integrates adversarial training into a variational inference framework. SAFER employs a two-stage optimization mechanism: first, deriving non-parametric policy distributions that satisfy safety constraints via convex optimization; and second, introducing adversarial state perturbations during supervised learning to enforce action distribution stability. This paradigm shift removes the dependency on online adversarial strategies, achieving robustness improvements using only benign offline data, thus avoiding safety hazards inherent in traditional methods.

In Goal-Conditioned RL (GCRL), the sparse binary reward mechanism limits the effectiveness of conventional adversarial training. To overcome this, Yin et al.\cite{yin2023rerogcrl} proposed Adversarially Robust Training with Semi-Contrastive Adversarial Augmentation (SCAA) and Sensitivity-Aware Regularization (SAR), collectively termed ARTs. SCAA maximizes the representational divergence between clean and perturbed states, effectively modeling perception-induced policy shifts while avoiding reliance on critic networks. SAR compensates for the lack of informative gradients under sparse rewards. By augmenting representation-based adversarial datasets, ARTs improve robustness in GCRL under conditions where traditional defense strategies fail.

In software-defined networking routing optimization, Altamirano et al.\cite{altamirano2024routing} addressed efficiency bottlenecks of adversarial training by proposing a distributed DRL framework integrated with Generative Adversarial Networks (GANs). The architecture assigns independent agents to each traffic class, achieving joint path optimization through single-action multi-node operations. The adversarial training mechanism allows the generator to learn dynamic network features, significantly reducing agent–network interactions and demonstrating the feasibility of adversarial training in large-scale network optimization. Similarly, Randhawa et al.\cite{randhawa2024deep}  introduced RELEVAGAN, a DRL-based adversarial evasion GAN designed for botnet detection under small-sample scenarios. The framework integrates DRL agents with an improved Auxiliary Classifier GAN (ACGAN), enabling the generation of semantically valid adversarial samples while maintaining detection evasion capabilities. Through a closed-loop adversarial training process, the discriminator simultaneously learns from balanced synthetic data and adversarial samples, eliminating the need for separate adversarial training phases.

Loevenich et al.\cite{loevenich2024drl} proposed a hybrid intelligent model that integrates Graph Neural Networks (GNNs) with DRL, aiming to significantly enhance the Quality of Service (QoS) optimization in tactical Mobile Ad Hoc Networks (MANETs) under dynamic adversarial environments. In this architecture, adversarial training is realized by constructing a state-adversarial MDP and injecting adversarial perturbations into the agent’s state observations through multi-stage attack strategies. By altering the transition dynamics of the original MDP, this approach forces the agent to explore suboptimal policy trajectories. The adversarial training mechanism adopts a Nash equilibrium-driven generative adversarial paradigm, enabling DRL agents to internalize adversarial feature patterns during policy optimization. Consequently, the framework improves robustness against non-stationary conditions such as network topology variations and link quality fluctuations.

In autonomous air traffic control, Wang et al.\cite{wang2025enhancing} introduced a Safety-Aware Deep Q-Network (SafeDQN) framework that decouples safety and efficiency objectives, improving interpretability. During adversarial training, they proposed a timed attack strategy, which selectively perturbs the model at specific time steps rather than continuously. This approach preserves attack effectiveness while significantly reducing perturbation frequency, yielding performance comparable to frequent attacks with reduced interference. The results highlight the role of adversarial training in simultaneously strengthening both safety and efficiency in DRL systems.

\subsection{Competitive Training}
Competitive training, as a game-theoretic co-optimization framework, enhances the robustness of RL policies by constructing dynamic adversarial environments where adversarial perturbation sources and agent policies form an evolving game relationship. This mechanism induces agents to explore non-stationary Nash equilibria in the policy space, thereby mitigating policy degradation caused by the violation of the Markov stationarity assumption in conventional single-agent training. Pinto et al.\cite{pinto2017robust} formalized RL problems as zero-sum games, requiring the protagonist agent to learn robust optimal policies under adversarial perturbations introduced by opponent agents. Compared to traditional RL paradigms, this framework effectively overcomes strategy degradation due to stationarity violations and significantly improves generalization in complex task settings, guiding agents toward more robust policy spaces. Abdullah et al.\cite{abdullah2019wasserstein} proposed the Wasserstein Robust RL (WR2L) framework, which introduces a competitive training paradigm based on Min-Max games to address policy generalization under dynamic environments. Specifically, WR2L models the policy optimization problem as a bi-level game between the policy network and an adversarial dynamic model. While the policy network seeks optimal decisions by maximizing expected cumulative returns, the adversarial model generates worst-case perturbation environments under a Wasserstein constraint to minimize policy performance. Through alternating optimization of policy and dynamic parameters, the competitive training process enables policies to gradually adapt to shifting distribution boundaries, effectively alleviating the overfitting issue of RL in fixed training environments.

In smart grid applications such as voltage control and power allocation, competitive training has been shown to alleviate sampling bias and local convergence issues inherent in single-agent DRL training. Veith et al.\cite{veith2024play} introduced a dual-agent adversarial training framework for voltage control, where grid operators and adversarial attackers form a dynamic game via inverse reward functions. This adversarial interaction compels both agents to continuously explore non-stationary boundary conditions, breaking through the policy degradation bottleneck caused by uniform sampling assumptions.

In the cooperative energy management of Virtual Power Plants (VPPs) aggregating multiple Microgrids (MGs)\cite{li2025deep}, competitive training constructs a dynamic game mechanism to reconcile conflicts between operators and microgrids, achieving Pareto-optimal solutions under multi-agent game equilibria. Specifically, VPP and MG agents form a bidirectional feedback loop through adversarial interactions, where competitive training guides strategy optimization via policy gradients. This ensures that while each agent maximizes its own payoff, pre-defined internal electricity price constraints are satisfied. Through dynamic adaptation, competitive training enables VPPs to achieve intelligent scheduling and resource allocation in volatile market environments, thereby improving system robustness and overall efficiency.

In multi-robot swarm control under dynamic adversarial conditions, Jia et al.\cite{jia2025deep} proposed an asymmetric self-play framework that introduces a learnable adversarial disturber to construct competitive training environments and stimulate the intelligent evolution of swarm strategies. The framework adopts a two-stage self-play paradigm: in the first stage, swarm robots and the disturber undergo synchronous co-evolution via adversarial optimization; in the second stage, a disturber model pool is established, and sampling weights are dynamically adjusted based on performance, forming a progressive curriculum learning mechanism. This hierarchical training approach effectively mitigates overfitting to single adversarial strategies and enhances policy generalization through diversified disturbance patterns.

\subsection{Robust Learning}
In the domain of adversarial defense and robustness research for DRL, robust learning focuses on enhancing policy fault tolerance and generalization by explicitly accounting for environmental uncertainties and potential adversarial factors during the training phase\cite{zhang2024reinforcement}. Unlike traditional approaches that reinforce models against fixed or single perturbations, robust learning typically incorporates multiple uncertainty assumptions or out-of-distribution scenarios\cite{mcallister2019robustness}, aiming to construct policy networks with stronger resistance to environmental changes, adversarial samples, and latent security threats. Behzadan et al.\cite{behzadan2018mitigation} proposed a parameter-space noise-driven robust learning framework, injecting adaptive noise into DQN weight parameters via the NoisyNet module to randomize exploration. This approach encourages the policy to explore a broader strategy space during training, thereby reducing overfitting to specific input patterns and mitigating cross-model transferability of adversarial examples. Addressing the vulnerability of DRL policies under model mismatch and adversarial attacks, Mandlekar et al.\cite{mandlekar2017adversarially} introduced the Adversarially Robust Policy Learning (ARPL) framework. ARPL actively constructs physically interpretable perturbations during training, introducing moderate system noise and parameter perturbations to improve policy generalization under real-world disturbances. It further employs white-box gradient information and curriculum learning to dynamically regulate perturbation signals, revealing the intrinsic relationship between policy robustness and system uncontrollability.

Traditional adversarial training faces high computational costs and limited generalization capabilities, making it insufficient for practical safety requirements\cite{yuan2019adversarial}. To overcome this, Yang et al.\cite{yang2023towards} proposed a data-centric robust learning paradigm (Data-centric Robust Learning, DRL), which builds deep learning models with strong robustness against transferable black-box adversarial attacks. This method innovatively generates highly transferable adversarial samples only once in the pre-training stage to construct a static augmented dataset and employs a confidence-driven data selection mechanism to optimize the sample distribution, breaking the optimization bottleneck of conventional adversarial training.

To address reward signal perturbations caused by sensor errors or adversarial attacks in real-world scenarios\cite{banihashem2021defense}, Wang et al.\cite{wang2020reinforcement} proposed a robust RL framework based on a reward confusion matrix. By constructing unbiased surrogate rewards, this method allows the agent to learn effectively even when only perturbed rewards are observed, without requiring any prior knowledge of the true reward distribution. Experiments in multiple Atari environments show significantly faster convergence and higher expected returns, with robust performance maintained even under high-noise conditions.

To tackle policy safety challenges arising from exploration uncertainty during deployment, Smirnova et al.\cite{smirnova2019distributionally} proposed a distributionally robust optimization framework with dynamic risk adjustment. By introducing a distributionally robust Bellman operator, the framework constructs a dynamic uncertainty set based on KL divergence during policy evaluation, minimizing worst-case value estimates to implement risk-averse exploration. This approach transforms adversarial policy optimization into a regularized sample reweighting process, providing a lower bound guarantee for state values while maintaining computational efficiency. The uncertainty set expands gradually as samples accumulate, ultimately converging to the optimal policy. Compared with traditional risk-sensitive methods, this approach avoids suboptimal policies caused by fixed risk biases and significantly improves early-stage robustness without sacrificing sample efficiency. Tessler et al.\cite{tessler2019action} proposed an Action Robust RL (AR-RL) framework to address execution deviations induced by noise, model uncertainty, or adversarial perturbations. Two novel action-robustness criteria were formalized: the Probabilistic Action Robust MDP (PR-MDP) and the Noisy Action Robust MDP (NR-MDP). PR-MDP introduces stochastic adversarial actions during training, simulating scenarios where the agent’s actions are partially perturbed, thereby enforcing fault-tolerant learning. NR-MDP models Gaussian noise in continuous action spaces and jointly optimizes perturbation bounds with policy updates to enhance stability under small noise. Both methods decouple perturbation generation from policy optimization in an adversarial game framework, dynamically balancing robustness and task performance while avoiding reliance on preset perturbation models. Oikarinen et al.\cite{oikarinen2021robust} proposed RADIAL-RL, a robust DRL framework based on adversarial loss, enhancing resilience to input perturbations, particularly under norm-constrained adversarial attacks. By leveraging neural network robustness verification theory, RADIAL-RL guides the policy network to maintain decision stability under perturbations, implicitly improving resistance to environmental uncertainties and malicious disturbances. The framework is compatible with various mainstream DRL algorithms and demonstrates superior training efficiency and robustness across benchmark tasks. Ren et al.\cite{ren2023promoting} developed a defense framework with temporal-awareness by integrating a dynamic critical-state detection module and a policy-space constraint mechanism. The detection module employs sliding-window analysis to monitor traffic flow anomalies in real time, identifying high-risk periods of potential adversarial perturbations through spatiotemporal correlation modeling. The policy-space constraint mechanism uses policy-gradient projection to limit deviations under adversarial conditions and combines historical policy trajectory distribution matching to maintain smoothness and stability in action selection. During training, adversarial traffic states are dynamically interpolated with normal samples in an adaptive ratio, and the defense policy is optimized to maximize the minimum performance bound on the mixed dataset, enhancing generalization to unseen attack patterns.

\subsection{Adversarial Detection}
Adversarial detection is a key defense strategy aimed at identifying and localizing anomalous behaviors induced by adversarial attacks. By monitoring the system's inputs and outputs in real time, detection mechanisms can determine whether a model is being maliciously perturbed, providing an effective means to safeguard robustness and security. Tekgul et al.\cite{tekgul2021real} innovatively proposed the Action Distribution Divergence Detector (AD3) for defending DRL policies. This method models the statistical properties of agent action sequences to detect adversarial perturbations. Based on the temporal coherence assumption of RL policies, AD3 uses Kullback-Leibler (KL) divergence to quantify the discrepancy between the conditional probability distribution of current actions and the reference distribution learned during pre-training. However, the method exhibits lower sensitivity to low-frequency or targeted attacks and may produce high false-positive rates in short-horizon tasks due to limited action samples.

Yang et al.\cite{yang2022training} proposed a causal inference framework that constructs a hierarchical defense system for adversarial detection. The framework employs a latent-space decoupling mechanism to separate high-frequency abnormal patterns of adversarial perturbations from environmental states and leverages adversarial contrastive learning to enhance the discriminative boundary of interference features. At the dynamic detection level, a temporal feature modeling module analyzes optical flow residuals and state continuity in observation sequences to capture spatiotemporal anomalies, such as trajectory disruptions caused by adversarial attacks. For detected perturbations, adaptive noise injection is applied in the feature space to disrupt the propagation path, and historical state information is fused to construct temporal self-healing capabilities. Experiments demonstrate that this detection system significantly improves the robustness threshold of models against gradient attacks and noise perturbations while maintaining decision sensitivity to benign observations. Xiang et al.\cite{xiang2018pca} addressed adversarial attacks in robotic path planning by proposing a principal component analysis (PCA)-based adversarial sample detection model. This method extracts principal features from normal path-planning data and assesses deviations from the normal distribution through projection errors to identify potential adversarial samples. It does not require prior knowledge of attack strategies, effectively overcoming weight allocation challenges faced by traditional detection methods. Experimental results show that the approach maintains high detection rates under varying degrees of adversarial perturbation, enhancing the robustness of Q-learning-based path-planning systems.

In the field of safe RL, adversarial detection has gradually shifted from passive constraint enforcement to active behavior modeling. Rahman et al.\cite{rahman2021adversarial} proposed the AdvEx-RL framework, which dynamically extracts unsafe behavior patterns through interaction with adversarial policies, establishing a task-independent safety firewall. By training optimal adversarial policies to maximize safety violation costs in the environment and evaluating state-action values dynamically, the framework implements a threshold-triggered post hoc safety shielding mechanism. Unlike traditional methods relying on manual constraints or offline dynamic modeling, AdvEx-RL employs maximum entropy RL to enhance exploration of adversarial strategies, uncover potential dangerous behaviors, and drive safe policy learning to maximize deviation from adversarial behavior distributions. This active identification and exclusion of unsafe actions establishes robust safety boundaries in dynamic and uncertain environments while avoiding performance degradation associated with joint optimization of task and safety objectives.

To address the vulnerability of traditional learning algorithms to catastrophic forgetting, Zhang et al.\cite{zhang2024memory} proposed an adversarial detection framework based on long-short-period memory. The framework analyzes historical data to identify and isolate non-cooperative behaviors. Long-term memory captures sustained deviations in separation, aggregation, and alignment objectives through a reinforcement-learning-driven cumulative credibility evaluation. Short-term memory employs dynamic neighborhood state evolution rules to filter out instantaneous anomalous behaviors. Periodic memory introduces frequency-domain analysis via fast Fourier transform to detect oscillatory strategies between cooperative and non-cooperative states. This multi-temporal detection paradigm effectively mitigates defense vulnerabilities caused by forgetting in traditional learning algorithms, providing a resilient and interpretable protective scheme for distributed multi-agent systems in complex adversarial environments.

\subsection{Defensive Distillation}
The concept of distillation was originally proposed by Hinton et al.\cite{hinton2015distilling}, with the core idea of transferring knowledge from a complex network to a simpler one. Papernot et al.\cite{papernot2016distillation} first introduced a defense-oriented distillation strategy in DNNs, aiming to enhance robustness against adversarial attacks by leveraging knowledge transfer and probability distribution softening mechanisms. In recent years, this concept has been extended to the domain of DRL\cite{chiejina2024system}, where the approach optimizes differences in the probability distributions of agent policies and introduces dynamic temperature adjustment, effectively improving stability and generalization under uncertain and adversarial conditions. Rusu et al.\cite{rusu2015policy} applied distillation to transfer policy knowledge learned by DRL agents to lightweight network architectures. By designing a supervised learning objective based on the Kullback-Leibler (KL) divergence, they addressed the challenges of distilling Q-value functions, which are prone to scale instability and action ambiguity, thereby achieving effective knowledge transfer from complex teacher networks to student networks. This approach also demonstrated strong policy integration capabilities in multi-task learning environments.

Traditional adversarial training methods rely on generating adversarial perturbations to enhance model robustness, but suffer from high computational cost and strong dependency on specific attack types. To address these limitations, Qu et al.\cite{qu2020defending} proposed a robust distillation loss under the defensive distillation paradigm, which does not require adversarial sample generation. The robust distillation loss consists of two components: Prescription Gap Maximization (PGM) loss and Jacobian Regularization (JR) loss. The PGM loss simultaneously optimizes the likelihood of teacher-selected actions and the entropy distribution of non-selected actions, encouraging the student policy to form significant action probability gaps under unperturbed conditions. The JR loss constrains the gradient magnitude of the loss function with respect to input states, reducing sensitivity to input perturbations and mitigating the gradient propagation effect of adversarial attacks. Theoretical analysis indicates that this composite loss can intrinsically enhance policy robustness under adversarial attacks without relying on specific attack generation mechanisms, by ensuring stable expansion of prescription gaps and the corresponding state-value differences.

Despite its effectiveness, research by Carlini et al.\cite{carlinib2016defensive} indicates that defensive distillation alone is insufficient to guarantee robust models, and must be combined with other mechanisms such as adversarial training and adversarial detection to construct a more resilient security framework. Czarnecki et al.\cite{czarnecki2019distilling} analyzed the limitations of defensive distillation in adversarial RL, noting that its offline training paradigm struggles to adapt to dynamic online attacks, prior dependence on specific attack models limits generalization, and the smoothing of data-space perturbations cannot address cumulative value bias in the decision space. Haydari et al.\cite{haydari2021adversarial} further pointed out that defensive distillation may fail to provide adequate protection against highly adaptive attack strategies and can be circumvented by advanced gradient-avoidance techniques. Compared with adversarial training, defensive distillation may still leave vulnerabilities when dealing with non-gradient-based attacks.

\subsection{Discussion}
This section categorizes and discusses adversarial defense methods in DRL, classifying them into five groups: adversarial training, competitive training, robust learning, adversarial detection, and defensive distillation. Table 2 systematically summarizes representative studies for each category. Existing research indicates that most defense mechanisms are primarily optimized for specific types of adversarial attacks and cannot comprehensively guard against all known or unknown threats. Each method has its strengths but also exhibits limitations, necessitating the selection of the most suitable strategy according to specific application scenarios and threat models.

Adversarial training enhances model robustness by incorporating adversarial examples but suffers from high computational costs and limited generalization. Competitive training, grounded in game theory, introduces adversarial agents to create dynamic game environments, improving policy resilience; however, it faces significant optimization challenges and convergence issues in high-dimensional tasks. Robust learning aims to improve fault tolerance from a policy optimization perspective, often leveraging distributional robustness or parameter noise to enhance adaptability to observational perturbations. Adversarial detection provides additional security guarantees but may struggle to accurately identify highly stealthy attacks. Defensive distillation reduces model sensitivity to small perturbations via knowledge transfer, yet it is less effective against high-intensity attacks.

Overall, current DRL defense strategies have not yet yielded a universal solution. Future research should focus on improving robustness while reducing computational overhead, integrating techniques such as causal reasoning, adaptive game-theoretic approaches, and distributionally robust optimization. The goal is to develop efficient and adaptable defense frameworks that balance security, computational resources, and task performance.

\begin{table*}[H]
  \centering
Table 2. Summary Of Defenses Against Adversarial Attacks On DRL
  \resizebox{\linewidth}{!}{
  \begin{tabular}{|c|m{4cm}<\centering|m{4cm}<\centering|m{4cm}<\centering|m{4cm}<\centering|}
  \hline
  \multicolumn{5}{|c|}{Competitive Training}\\
    \hline
    \multirow{2}{*}{Paper} & \multirow{2}{*}{Proposed Methods} & \multicolumn{2}{c|}{Setup} & \multirow{2}{*}{Effective Against} \\
    \cline{3-4}
    & & Algorithm & Environment & \\
    \hline
    Pinto et al. {[}73{]} & Robust Adversarial Reinforcement Learning & TRPO & InvertedPendulum, HalfCheetah, Swimmer, Hopper, Walker2d & Environmental Disturbance Attacks \\
    \hline
    Abdullah et al. {[}74{]} & Wasserstein Robust Reinforcement Learning & PPO, TRPO, RARL & CartPole, Hopper, Walker2D, Halfcheetah & Model Perturbation Attacks \\
    \hline
    Veith et al. {[}75{]} & Autocurriculum Training for Adversarial Learning & SAC & Power Grid & System Perturbation Attacks \\
    \hline
    Li et al. {[}76{]} & DRL-based Hierarchical Energy Management for VPP & PPO & Virtual Power Plant with multiple Microgrids & Pricing Strategy Attacks \\
    \hline
    Jia et al. {[}77{]} & Asymmetric Self-Play Flocking Control & PPO & Multi-robot flocking, dynamic obstacles and interference & Dynamic Obstacle Attacks \\
    \hline
    \multicolumn{5}{|c|}{Adversarial training}\\ \hline
    \multirow{2}{*}{Paper} & \multirow{2}{*}{Proposed Methods} & \multicolumn{2}{c|}{Setup} & \multirow{2}{*}{Effective Against} \\
    \cline{3-4}
    & & Algorithm & Environment & \\
    \hline
    Kos et al. {[}66{]} & Adversarial Training using FGSM and Random Noise & A3C & Pong & FGSM \& Random Noise Attack \\
    \hline
    Liu et al. {[}67{]} & Adversarial Training using SAFER & ADV-PPOL, SA-PPOL, CVPO & Ball-Circle, Car-Circle, Drone-Run, Ant-Run & Observation Perturbations \\
    \hline
    Yin et al. {[}68{]} & Adversarial Training using ARTs & DDPG, GCSL, GoFar & FetchPush, FetchPick, FetchReach, FetchSlide & Observation \& Goal Perturbations \\
    \hline
    Altamirano et al. {[}69{]} & DRL-based Routing with GAN-enhanced Training & DDQN & Containernet & Sparse Observation Perturbations \\
    \hline
    Randhawa et al. {[}70{]} & Adversarial training using GAN & DDQN & Botnet Detection & Adversarial Evasion Attacks \\
    \hline
    Loevenich et al. {[}71{]} & GNN-assisted DRL with Adversarial Training for Tactical QoS & PPO, A2C, TRPO, DQN, RecurrentPPO, ARS & Tactical MANETs & Observation Perturbations \\
    \hline
    Wang et al. {[}72{]} & Adversarial training with SafeDQN using goal/safety-split Q-networks & DQN, SafeDQN & Air Traffic Control & Observation Perturbations \\
    \hline
     \multicolumn{5}{|c|}{Robust Learning}\\ \hline 
       \multirow{2}{*}{Paper} & \multirow{2}{*}{Proposed Methods} & \multicolumn{2}{c|}{Setup} & \multirow{2}{*}{Effective Against} \\
    \cline{3-4}
    & & Algorithm & Environment & \\
    \hline
    Behzadan et al. {[}80{]} & Parameter-Space Noise for Mitigation of Policy Manipulation Attacks & DQN & Enduro, Assault, Breakout & Policy Manipulation Attacks \\
    \hline
    Mandlekar et al. {[}81{]} & Adversarially Robust Policy Learning & TRPO & Inverted Pendulum, HalfCheetah, Hopper, Walker & State Perturbation Attacks \\
    \hline
    Yang et al. {[}83{]} & Data-centric Robust Learning & PGD, PGD-AT, TRADES, EAT, FAT & CIFAR-10, CIFAR-100, TinyImageNet & Transfer-based Adversarial Attacks \\
    \hline
    Wang et al. {[}85{]} & Robust RL with Perturbed Rewards & Q-Learning, DQN, PPO & CartPole, Pendulum, Atari Games & Reward Perturbation Attacks \\
    \hline
    Smirnova et al. {[}86{]} & Distributionally Robust Policy Iteration & DR-SAC & Hopper, Walker2D & Environmental Perturbation Attack \\
    \hline
    Tessler et al. {[}87{]} & Action Robust Reinforcement Learning & PR-MDP, NR-MDP, AR-DDPG & Hopper, Walker2D, Humanoid, InvertedPendulum & Action Perturbation Attacks \\
    \hline
    Oikarinen et al. {[}88{]} & Robust Adversarial Loss & DQN, A3C, PPO & Atari games, MuJoCo, ProcGen & $l_p$-norm Attacks \\
    \hline
    Ren et al. {[}89{]} & Adversarial Robustness and Reward Modifications & MADDPG & SUMO multi-intersections & Reward Poisoning Attacks \\
    \hline
  \end{tabular}}
  \label{tab:competitive_training}
\end{table*}

\begin{table*}[h]
  \centering 
  (Continued) Table 2. Summary Of Defenses Against Adversarial Attacks On DRL
  \resizebox{\linewidth}{!}{
  \begin{tabular}{|c|m{4cm}<\centering|m{4cm}<\centering|m{4cm}<\centering|m{4cm}<\centering|}
  \hline
  \multicolumn{5}{|c|}{Adversarial Detection}\\
  \hline
    \multirow{2}{*}{Paper} & \multirow{2}{*}{Proposed Methods} & \multicolumn{2}{c|}{Setup} & \multirow{2}{*}{Effective Against} \\
    \cline{3-4}
    & & Algorithm & Environment & \\
    \hline
    Tekgul et al. {[}90{]} & Adversarial Detection using Universal Adversarial Perturbations & DQN, PPO, A2C & Pong, Breakout, Freeway & Universal Adversarial Perturbations \\
    \hline
    Yang et al. {[}91{]} & Causal Inference Q-Network for training & DQN & Cartpole, LunarLander, Banana Collector, Pixel Cartpole & Observation Perturbations \\
    \hline
    Xiang et al. {[}92{]} & PCA-Based Model for Adversarial Detection & Q-learning & Grid-world with random obstacles & Pathfinding Attacks \\
    \hline
    Rahman et al. {[}93{]} & Adversarial Behavior Exclusion & SAC, DQN, PPO & MuJoCo, SafetyGym & Safety Violations \\
    \hline
    Zhang et al. {[}94{]} & Memory-based adversarial detection & Q-learning & Multi-Agent Flocking in Dynamic Environments & Adversarial Agent Behaviors, Policy Manipulation Attacks \\
    \hline
  \multicolumn{5}{|c|}{Defensive Distillation}\\
  \hline 
    \multirow{2}{*}{Paper} & \multirow{2}{*}{Proposed Methods} & \multicolumn{2}{c|}{Setup} & \multirow{2}{*}{Effective Against} \\
    \cline{3-4}
    & & Algorithm & Environment & \\
    \hline
    Rusu et al. {[}98{]} & Policy Distillation for reinforcement learning & DQN & Atari Games (Breakout, Pong, etc.) & Gradient-based Attacks, Model Overfitting \\
    \hline
    Qu et al. {[}99{]} & Defensive Distillation & DDQN, Rainbow DQN & Pong, Boxing, Freeway & Generic Adversarial Attacks \\
    \hline
    Czarnecki et al. {[}101{]} & Policy distillation & Q-Learning, Actor-Critic, Policy Gradient & Grid-world & Generic Adversarial Attacks \\
    \hline
  \end{tabular}}
  \label{tab:competitive_training}
\end{table*}

\label{ToADoD}
\section{Current issues and Prospects}
Despite the remarkable progress of DRL in a wide range of domains, ensuring its security under adversarial environments remains highly challenging. Existing adversarial attack and defense techniques have enhanced the robustness of agents to a certain extent; however, they still suffer from notable limitations in terms of generalization ability, computational efficiency, scalability, interpretability, comprehensiveness of evaluation frameworks, and hardware-level security. These deficiencies significantly hinder the widespread adoption and reliable deployment of DRL in safety-critical applications. This section highlights the key issues in adversarial security research of DRL, systematically analyzes the major challenges DRL faces when countering adversarial attacks and security threats, and discusses promising directions for future research.

\subsection{Generalization}
Existing defense strategies in DRL are largely tailored to specific types of adversarial attacks. A common approach, adversarial training, improves model robustness by incorporating known attack samples. Nevertheless, such methods often fail to generalize to unseen or mutated attacks, limiting the agent’s adaptability across diverse environments and tasks\cite{moosavi2017universal}. Furthermore, many proposed defenses demonstrate effectiveness only under controlled experimental conditions, yet lack transferability to real-world scenarios\cite{ren2021adversarial}. In highly dynamic environments, attack patterns may evolve over time, rendering static training-based defenses ineffective in the long run\cite{cao2024deep}. Striking a balance between adversarial robustness, generalization capability, and deployment efficiency thus remains a central challenge in DRL security. Future research is expected to emphasize adaptive and cross-environment generalization of defense strategies while reducing the computational overhead in practical applications. Leveraging approaches such as meta-learning\cite{lotfi2024meta} and transfer learning\cite{janiar2024intelligent}, DRL defenses may become more resilient against previously unseen attacks, thereby better addressing the challenges posed by dynamic and evolving environments.

\subsection{Computational Complexity}
Enhancing the security of DRL often demands substantial additional computational resources, and the incorporation of adversarial defense mechanisms further exacerbates this burden.\cite{buivh2024critical} In particular, within high-dimensional state spaces, adversarial training requires the generation of large volumes of adversarial samples and continuous robustness evaluation during policy optimization, which significantly increases both training and inference costs. Moreover, real-time attack detection necessitates continuous monitoring of input data and rapid identification of potential adversarial perturbations, imposing even stricter demands on computational resources. In high-security applications, DRL agents must make critical decisions within extremely short time windows, yet existing defense mechanisms often struggle to deliver stringent security guarantees while maintaining efficient inference\cite{zhang2023multi}. The challenge is further amplified in resource-constrained environments, such as embedded systems and edge computing, where limited computational capacity and energy efficiency hinder the direct deployment of computation-intensive adversarial training and real-time detection\cite{zhang2024advancing}. Consequently, striking a balance among computational efficiency, response latency, and security has become a central challenge in DRL adversarial safety research. Looking ahead, targeted optimization of resource allocation, the development of lightweight defense mechanisms, and the integration of efficient parallel computing techniques are expected to be key directions for enhancing the security of DRL systems.

\subsection{Scalability}
In complex environments, the security of DRL depends not only on the agent’s decision-making capability but also on the dynamics of the environment and the interactions among multiple agents. Traditional safe RL methods perform well in fixed environments; however, their adaptability and scalability remain severely challenged in dynamic, non-stationary, or high-dimensional settings. On one hand, as environmental complexity increases, the dimensionality of the state and action spaces grows rapidly, substantially raising the difficulty of training and computational costs. As a result, existing defense strategies often fail to remain effective in larger-scale or more complex environments. On the other hand, real-world environments are inherently unpredictable, and the data distribution on which DRL agents are trained may deviate from deployment conditions. This distributional shift can lead to erroneous decisions in unseen or abnormal situations, leaving the agent vulnerable to adversarial attacks\cite{qian2024offline}. Moreover, RL models typically assume the Markov property of environments, whereas real-world systems frequently exhibit non-Markovian characteristics, further complicating the design of effective defense strategies\cite{rupprecht2022survey}. Thus, improving both the scalability and adaptability of DRL while ensuring security is a pressing research challenge. Future directions will likely focus on enhancing policy generalization, strengthening online adaptability, and developing adaptive defense mechanisms, enabling DRL agents to operate reliably and securely despite environmental changes and evolving adversarial threats.

\subsection{Explainability}
The “black-box” nature of DRL models makes their decision-making processes difficult to interpret[17]. This lack of transparency not only undermines the credibility of adversarial defense mechanisms but also hinders their deployment in domains with stringent security requirements. For instance, in applications such as smart grids\cite{zhang2018review} and medical diagnosis\cite{yu2021reinforcement}, the safety-critical decisions made by DRL agents must be sufficiently transparent to allow system operators to understand the underlying reasoning and conduct effective risk assessments. However, most existing research on adversarial attacks and defenses has primarily focused on improving robustness, with comparatively little attention paid to interpretability and trustworthiness. Moreover, the absence of intuitive visualization tools and systematic risk assessment frameworks makes it difficult for agents to detect potential threats or respond effectively to adaptive attacks in complex environments. Therefore, a key direction for future research lies in developing interpretable adversarial analysis and defense strategies, enabling DRL agents to provide understandable decision rationales when confronted with adversarial perturbations.

\subsection{Evaluation Metrics}
Although numerous adversarial attack and defense methods have emerged in recent years, a unified evaluation standard for comprehensively assessing the security and robustness of DRL under attack is still lacking\cite{ilahi2021challenges}. Most studies rely on task-specific settings and particular attack scenarios, without conducting systematic validation across broader environments and diverse attack types. Moreover, current evaluation practices are often centered on performance metrics, while paying insufficient attention to critical factors such as safety constraints, computational cost, and real-time requirements. This limitation not only hampers a holistic understanding of the impact of adversarial attacks but also restricts fair comparison among different defense strategies. Therefore, establishing a standardized evaluation framework that incorporates diverse attack modalities, task environments, and resource constraints is an essential step toward advancing DRL adversarial safety and enabling equitable benchmarking of attack and defense approaches.

\subsection{Hardware Security}
Beyond algorithmic attacks, DRL systems also face hardware-level security threats in practical applications\cite{liang2020deep}. For example, sensor spoofing attacks can mislead agents by feeding them falsified environmental information, thereby inducing erroneous decisions\cite{parras2021deep}. In addition, within edge or cloud computing environments, adversaries may exploit adversarial techniques to manipulate computational resources, tamper with data transmission, or steal model parameters, ultimately compromising the inference process of DRL systems. Such physical-layer security threats are often difficult to fully mitigate using traditional software-based defenses alone. Consequently, future research should explore integrated hardware–software defense strategies, such as trusted execution environments\cite{zhang2025trust}, hardware-based cryptographic modules\cite{guan2022overview}, and robust sensing technologies\cite{jiang2024ris}, to enhance the security and stability of DRL in complex physical environments.

\label{CIaP}
\section{Conclusion}
As a key research direction in autonomous learning, DRL has achieved remarkable progress in areas such as strategy games, multi-agent systems, autonomous driving, and mobile robotics. With the expansion of its application scope, concerns regarding security and robustness have increasingly emerged, particularly in the context of adversarial attacks. Enhancing the resilience of DRL agents against such perturbations has become a critical challenge. This paper provides a systematic review of recent advances in adversarial attacks and defense strategies for DRL, proposes a classification framework for adversarial attacks based on perturbation types and attack targets, and summarizes adversarial training methods aimed at improving agent robustness. In addition, we discuss several open challenges in current research, including limited generalization, high computational complexity, poor scalability, insufficient interpretability, and the lack of standardized evaluation frameworks. Future research is expected to focus on enhancing the stability, adaptability, and decision reliability of DRL agents in complex and dynamic adversarial environments, thereby promoting the safe deployment of DRL technologies across a broader range of applications.
\label{Conclusion}

\printcredits

\bibliographystyle{elsarticle-num-names} 

\bibliography{cas-refs}



\end{document}